\documentclass[english, reprint, aps, prb, twocolumn, superscriptaddress, showpacs, floatfix]{revtex4-1}
\usepackage{babel}
\usepackage{amssymb}

\usepackage{amsmath}
\usepackage{float}
\usepackage{bbm}
\usepackage{graphicx}
\usepackage{epsfig}
\usepackage{verbatim}
\usepackage[usenames]{color}

\usepackage{hyperref}
\begin{document}

\title{Robust quantum gates for singlet-triplet spin qubits using composite pulses}
\author{Xin Wang}
\affiliation{Condensed Matter Theory Center, Department of Physics, University of Maryland, College Park, MD 20742, USA}
\author{Lev S. Bishop}
\affiliation{Condensed Matter Theory Center, Department of Physics, University of Maryland, College Park, MD 20742, USA}
\affiliation{Joint Quantum Institute, University of Maryland, College Park, MD 20742, USA}
\author{Edwin Barnes}
\affiliation{Condensed Matter Theory Center, Department of Physics, University of Maryland, College Park, MD 20742, USA}
\affiliation{Joint Quantum Institute, University of Maryland, College Park, MD 20742, USA}
\author{J.~P.~Kestner}
\affiliation{Department of Physics, University of Maryland Baltimore County, Baltimore, MD 21250, USA}
\affiliation{Condensed Matter Theory Center, Department of Physics, University of Maryland, College Park, MD 20742, USA}
\author{S.~Das Sarma}
\affiliation{Condensed Matter Theory Center, Department of Physics, University of Maryland, College Park, MD 20742, USA}
\affiliation{Joint Quantum Institute, University of Maryland, College Park, MD 20742, USA}
\date{\today}

\begin{abstract}
We present a comprehensive theoretical treatment of {\sc supcode}, a method for generating dynamically corrected quantum gate operations, which are immune to random noise in the environment, by using carefully designed sequences of soft pulses. {\sc supcode} enables dynamical error suppression even when the control field is constrained to be positive and uniaxial, making it particularly suited to counteracting the effects of noise in systems subject to these constraints such as singlet-triplet qubits. We describe and explain in detail how to generate {\sc supcode} pulse sequences for arbitrary single-qubit gates and provide several explicit examples of sequences that implement commonly used gates, including the single-qubit Clifford gates. We develop sequences for noise-resistant two-qubit gates for two exchanged-coupled singlet-triplet qubits by cascading robust single-qubit gates, leading to a 35\% reduction in gate time compared to previous works. This cascade approach can be scaled up to produce gates for an arbitrary-length spin qubit array, and is thus relevant to scalable quantum computing architectures. To more accurately describe real spin qubit experiments, we show how to design sequences that incorporate additional features and practical constraints such as sample-specific charge noise models and finite pulse rise times. We provide a detailed analysis based on randomized benchmarking to show how {\sc supcode} gates perform under realistic $1/f^\alpha$ noise and find a strong dependence of gate fidelity on the exponent $\alpha$, with best performance for $\alpha>1$. Our {\sc supcode} sequences can therefore be used to implement robust universal quantum computation while accommodating the fundamental constraints and experimental realities of singlet-triplet qubits. 
\end{abstract}

\pacs{03.67.Pp, 03.67.Lx, 73.21.La}

\maketitle

\section{introduction}

A quantum computer would possess the fascinating ability to perform certain computational tasks exponentially faster than classical computers, by nontrivially using the exponentially large size of a many-body quantum Hilbert space. \cite{NielsenChuang.00} Semiconductor quantum dot spin systems are one of the leading candidates for building a quantum computer because of their prospective scalability,\cite{Taylor.05} their long coherence times,\cite{Bluhm.10b} and their capacity for fast all-electrical gate operations.\cite{Petta.05, Maune.12} There are various ways to encode quantum information in the spin states of electrons loaded into one or more quantum dots. For example, the two spin states of a single electron can form a qubit;\cite{Loss.98} alternatively a qubit may also be encoded in the collective spin states of two\cite{Levy.02} or three electrons.\cite{DiVincenzo.00,Laird.10,Medford.13} In this paper, we focus on the case of the singlet-triplet qubit,\cite{Petta.05} where the qubit is encoded in the singlet-triplet spin subspace of two electrons trapped in a double quantum dot. This encoding scheme has the advantages of fast single-qubit operations and of being immune to homogenous fluctuations of the magnetic field. Arbitrary single-qubit operations are performed by combining $z$-axis rotations around the Bloch sphere, achieved by a tunable exchange interaction between the singlet and triplet states,\cite{Petta.05} and $x$-axis rotations, which are generated by a local magnetic field gradient.\cite{Foletti.09,Bluhm.10a,Brunner.11,Petersen.13} Together with an entangling two-qubit gate, which can be based on either a capacitive coupling\cite{Shulman.12} between the two qubits or an exchange coupling,\cite{Klinovaja.12} one is then able to perform universal quantum computation. The great advantage of the singlet-triplet quantum dot spin qubits, leading to substantial experimental and theoretical activities in the topic, is that the qubit operations can all be implemented by external electric fields (i.e. suitable gate voltages), thus making them operationally convenient as well as compatible with existing semiconductor electronics.

One of the biggest obstacles to the realization of a quantum computer is the qubit decoherence that results from the interaction between the qubits and their environment. This decoherence must be very small for successful quantum computation to work, and the central problem of the whole field has been the issue of whether it is experimentally  feasible to reduce decoherence to a level low enough for fault-tolerant quantum computation to go forward---in particular, the decoherence must be very small both during the idling of the gates (i.e. when the qubits are just quantum memory) and during the actual gate operations. There are two main noise channels for singlet-triplet qubits leading to decoherence: Overhauser noise, which stems from the hyperfine-mediated spin flip-flop processes that take place between the electron spins and the nuclear spins in the surrounding substrate,\cite{Reilly.08, Cywinski.09,Barnes.12} and charge noise arising from environmental voltage fluctuation, which corresponds to the deformation of the quantum dot confinement potential due to nearby impurities or other sources of uncontrolled stray electric fields.\cite{Hu.06, Culcer.09, Nguyen.11} Fortunately, these types of noise are highly non-Markovian: they produce stochastic errors in the qubit Hamiltonian which vary on a much longer time scale ($\sim 100\mu$s) than typical gate operation times (on the scale of ns). Dynamical decoupling has proven to be a successful method for combating this kind of noise. Its underlying idea is the ``self-compensation'' of errors, best illustrated by the Hahn spin echo technique introduced first in the context of NMR:\cite{Hahn.50} when a quantum state dephases due to noise over some time span, one may apply a $\pi$-pulse to flip the sign of the error in the state, effectively reversing the error's evolution so that the qubit ``refocuses'' to its original state after a second time span of equal duration to the first. Here it is very important that the noise is non-Markovian since one requires the noise to remain static over the time spans before and after the $\pi$-pulse. This dynamical way of reviving a quantum state has proven invaluable to coherent manipulation of quantum systems, as have several more sophisticated pulse sequences that were subsequently developed\cite{Carr.54,Meiboom.58,Khodjasteh.05,Uhrig.07,Green.13} and implemented in experiments.\cite{Barthel.10,Bluhm.10b,Medford.12} In general, dynamical decoupling extends the coherence time from the dephasing time $T_2^*$ to a much longer timescale $T_2$ (which is defined depending on the specific dynamical decoupling sequence used) beyond which the quantum information is inevitably lost. For singlet-triplet spin qubits in GaAs quantum dots, $T_2^*\sim10$ ns and $T_2\sim0.1$ ms,\cite{Petta.05,Bluhm.10b} while for Si $T_2^*\sim100$ ns\cite{Maune.12} and $T_2\sim0.1$ ms\cite{Pla.12} but is expected to be even longer in isotope-enriched samples.\cite{Witzel.10,Tyryshkin.12} Therefore, dynamical decoupling is a powerful way to preserve a quantum state against noise, enabling robust quantum memory.

Achieving robust quantum memory capabilities, however, covers only one of the requirements for a viable quantum computer. Equally necessary is the ability to protect the qubit from noise {\it while} performing quantum gates on it. This necessity has motivated the development of dynamically corrected gates (DCGs),\cite{Goelman.89,Khodjasteh.09,Khodjasteh.10,Bensky.10,Grace.12,Green.12,Kosut.13} which can roughly be thought of as an extension of dynamical decoupling to the situation where the qubit is simultaneously being purposefully rotated. In particular, DCGs also typically exploit the notion of self-canceling errors. Like dynamical decoupling, such protocols have been vastly successful in NMR and in the general theory of quantum control. However, in contrast to dynamical decoupling, most approaches to DCGs developed thus far in the literature are not applicable to the case of singlet-triplet qubits because of their unique experimental constraints. First, the tunable exchange interaction which gives rise to $z$-axis rotations is always non-negative and bounded from above by a certain maximal value.\cite{Petta.05,Maune.12} Second, in order to do arbitrary single-qubit rotations, one must set up a magnetic field gradient\cite{Foletti.09,Barthel.10} across the two quantum dots; this gradient cannot be varied during gate operations, meaning that the control is effectively single-axis (along $z$) and that there is an always-on field rotating the qubit states into each other. Either constraint by itself would already rule out many DCG schemes; together, these constraints make noise-resistant control in singlet-triplet qubits uniquely challenging. In particular, the spectacular pulse control techniques developed in the NMR literature over many years are useless for our purpose since NMR does not satisfy the special constraints discussed above, and we must start from scratch and develop DCG pulses for the singlet-triplet qubits obeying the special constraints of the problem.

Despite these challenges, it was realized recently that it is still possible to develop DCGs for singlet-triplet qubits subject to static noise. In Ref.~\onlinecite{Wang.12}, we introduced {\sc supcode} (Soft Uniaxial Positive Control for Orthogonal Drift Error), demonstrating that it is possible to design special sequences of square pulses that implement robust quantum gates while at the same time respecting all experimental constraints. {\sc supcode} was originally introduced to cancel errors due to Overhauser noise only. In the case of a non-zero magnetic field gradient, we showed how to cancel the leading-order effect of Overhauser noise by supplementing a na\"ive pulse with an uncorrected identity operation, designed in such a way that the errors accumulated during the identity operation exactly cancel the errors arising during the na\"ive pulse. We further showed that by performing the identity operations as interrupted $2\pi$ rotations around certain axes of the Bloch sphere, error cancelation is always possible since one has the flexibility to include as many degrees of freedom as necessary for the cancelation simply by including more interruptions. The cost one has to pay is that the error-correcting pulse is typically substantially longer than the na\"ive pulse. For the cases discussed in Ref.~\onlinecite{Wang.12}, more than $40\pi$ of rotation around the Bloch sphere is required for an error-correcting pulse. A long pulse sequence is an essential price to pay for carrying out error-corrected DCG operations in quantum computation, but the pulse time can be optimized through careful calculations.

This idea of correcting a na\"ive pulse by supplementing it with an identity operation formed by nested $2\pi$ rotations was further developed and optimized in Ref.~\onlinecite{Kestner.13}. There, we showed that arbitrary single-qubit rotations can be made resistant to both Overhauser and charge noise simultaneously. Furthermore, it was shown that the pulse sequence duration can be reduced by a factor of $\sim2$ from the previous work, Ref.~\onlinecite{Wang.12}, even though the sequences cancel both types of noise, not just Overhauser noise, greatly increasing the experimental feasibility of these sequences. Subsequently, alternative approaches to DCGs for canceling both types of
noise in singlet-triplet qubits have  appeared in the
literature.\cite{Khodjasteh.12}

In Ref.~\onlinecite{Kestner.13}, we also showed that {\sc supcode} can be extended to construct robust two-qubit exchange gates based on the inter-qubit exchange-coupling, and that it is again possible to protect against both Overhauser and charge noise. The design of a robust two-qubit gate is considerably more complicated because of the presence of additional errors that do not arise in the single-qubit case, including possible leakage error out of the computational subspace as well as the over-rotation error in the two-qubit Ising gate caused by charge noise. Nevertheless we have shown that these obstacles can be circumvented when single-qubit {\sc supcode} gates are combined in a manner similar to the BB1 sequence developed in NMR.\cite{Wimperis.94,Jones.03} Unfortunately, the resulting sequence is relatively long (about $360\pi$ of rotation) and is challenging for actual implementation in the laboratory. The task then remains to reduce the length of the pulse sequence while maintaining its robustness against noise.

The main purpose of this paper is to bridge the gap between the theory of {\sc supcode} and its experimental implementation. As in the development of any theory, we have made several simplifying assumptions. First, it is generally the case that the qubit exchange coupling is controlled by the tilt, or detuning, of the double quantum dot confinement potential. This allows the experimenter to control the qubit by adjusting voltages, but it also makes the qubit vulnerable to charge noise. Furthermore, the effect of charge noise on the qubit will generally depend on the precise dependence of the exchange coupling on the detuning. In our previous works, we have mostly assumed a phenomenological relation between the exchange coupling $J$ and detuning $\epsilon$: $J(\epsilon)\propto\exp(\epsilon/\epsilon_0)$, a form used in previous works.\cite{Shulman.12,Dial.13} However, this phenomenological form is non-universal, and in practice $J(\epsilon)$ varies from sample to sample. It is therefore an important question to ask whether {\sc supcode} would still work for other charge noise models in which $J(\epsilon)$ has a different form. Second, we have assumed that the pulses are perfect square pulses which are turned on and off instantaneously. In actual experiments, the pulses have finite rise times, and in Ref.~\onlinecite{Wang.12}, we have shown that inclusion of the finite rise time would only amount to a shift in pulse parameters but otherwise leave our major results unchanged for the original {\sc supcode}. The question remains whether the same holds for the more powerful yet shorter sequences presented in Ref.~\onlinecite{Kestner.13}. In this paper, we explicitly examine these experimental considerations and show that the power of {\sc supcode} sequences is not compromised by the extra complications of real systems. We further clarify how one could slightly modify the pulse parameters of the two-qubit gate in order to accommodate different charge noise models. Moreover, we show that the length of the corrected two-qubit gates can be reduced by as much as 35\% from that shown in Ref.~\onlinecite{Kestner.13}, a significant step toward future experimental implementation. We believe that the optimized DCG pulse sequence proposed in the current article are ready for immediate implementation in the laboratory spin qubit experiments.

Most crucially, in the previous works we have assumed a static noise model. Such a model captures the essence of the quasi-static noise found in actual experiments,\cite{Medford.12} and the basic idea is that in such realistic situations, performing a {\sc supcode} sequence would echo away most, although not all, of the effect of the noise. In this paper, we test this idea by performing randomized benchmarking of the 24 single-qubit Clifford gates, all found through our {\sc supcode} framework, under $1/f^\alpha$ noise, where $\alpha$ is a parameter that depends on the physical processes causing the noise. We show that unlike for static noise, in this case there is a limit to the amount of improvement possible via \textsc{supcode}, but that this limit depends strongly on $\alpha$ and substantial benefit from \textsc{supcode} is available for the case where $\alpha>1$. The results we present in this paper show that {\sc supcode} is a powerful tool that can perform noise-resistant quantum gates despite the complications of real spin qubit systems, including different dependencies between the exchange coupling and the detuning, finite rise times and realistic $1/f^\alpha$ noise sources. For these reasons, we believe that {\sc supcode} will be immensely helpful to on-going experimental efforts in performing quantum gates on semiconductor quantum dot devices.

This paper is organized as follows. In Sec.~\ref{sec:model} we present the theoretical model, explain the experimental constraints and the basic assumptions that we have made. In Sec.~\ref{sec:oneqrot} we give a very detailed and pedagogical review of how {\sc supcode} sequences are constructed for an arbitrary single-qubit rotation. Explicit examples of several quantum gates are also presented, including the 24 single-qubit Clifford gates which are used in the randomized benchmarking in Sec.~\ref{sec:benchmarking}. We discuss how different charge noise models and finite rise times would affect our {\sc supcode} sequences in Sec.~\ref{sec:altgofJ} and Sec.~\ref{sec:finiterisetime}, respectively. In Sec.~\ref{sec:twoq} we show that the length of the corrected two-qubit gate presented in Ref.~\onlinecite{Kestner.13} can be significantly reduced in duration by about 35\%. We also show how the pulse parameters are minimally altered for a general charge noise model. Following this, we discuss the noise-resistant manipulation of a multi-qubit system using single-qubit and two-qubit corrected gates presented in this paper and the buffering identity operation required to accomplish this task. We present randomized benchmarking results in Sec.~\ref{sec:benchmarking}. Finally we conclude in Sec.~\ref{sec:conclusion}.

\section{Model and Basic assumptions}\label{sec:model}

The model Hamiltonian for a singlet-triplet qubit can be expressed in terms of the Pauli operators $\mathbf{\sigma}$ as
\begin{equation}\label{eq:ham}
H(t)=\frac{h}{2}\sigma_x+\frac{J\left[\epsilon\left(t\right)\right]}{2}\sigma_z.
\end{equation}
The computational bases are
$|0\rangle=|\mathrm{T}\rangle=({|\!\uparrow\downarrow\rangle}+{|\!\downarrow\uparrow\rangle})/\sqrt{2}$ and $|1\rangle=|\mathrm{S}\rangle=\left({|\!\uparrow\downarrow\rangle}-{|\!\downarrow\uparrow\rangle}\right)/\sqrt{2}$.
Here, ${|\!\downarrow\uparrow\rangle}=c_{1\downarrow}^\dagger c_{2\uparrow}^\dagger|\mathrm{vacuum}\rangle$, where $c_{j\sigma}^\dagger$ creates an electron with spin $\sigma$ at the $j$th dot. Any linear combinations of the $|0\rangle$ and $|1\rangle$ states can be represented as a unit vector pointing towards a specific point on the Bloch sphere, with $|0\rangle$ and $|1\rangle$ its north and south poles, respectively. Being able to perform arbitrary single qubit operations then amounts to being able to rotate such a unit vector---the Bloch vector---from any point to any other point on the Bloch sphere. This capability combined with an entangling two-qubit gate, such as the {\sc cnot} gate, suffices to achieve universal quantum computation.

Geometrically, one needs the ability to rotate around two non-parallel axes of the Bloch sphere in order to complete an arbitrary rotation. In this system,
rotations around the $x$-axis are performed with a magnetic field gradient across the double-dot system, which in energy units reads $h=g\mu_B\Delta B_z$. In practice the magnetic field gradient is generated either by dynamically polarizing the nuclear spins surrounding the double dots\cite{Foletti.09,Bluhm.10a} (the ``Overhauser field''),  or by depositing a permanent micromagnet nearby.\cite{Brunner.11,Petersen.13} In principle, the magnetic field gradient can be changed, and thus also the rotation rate around the $x$-axis. Unfortunately, changing it requires times much longer than the gate operation time. Therefore we assume that in performing a given computational task, the magnetic field gradient, $h$, is held constant throughout.

Rotations around the $z$-axis are done by virtue of the exchange interaction $J$, the energy level splitting between $|\mathrm{S}\rangle$ and $|\mathrm{T}\rangle$. A nice feature of the quantum dot system is that the magnitude of $J$ can be controlled by the detuning  $\epsilon$, namely the tilt of the effective double-well confinement potential, which in turn can be done by simply changing the gate voltages. In other words, by feeding in a series of carefully designed pulses to the control gates, one then has fast, all-electrical control of the rotation rate around the $z$-axis.  However, due to its intrinsic energy level structure,\cite{Petta.05} $J$ is bounded from below by zero, and from above by a certain maximal value $J_{\rm max}$, beyond which the tunneling between quantum dots becomes large enough to alter the charge configuration of the electrons. (In certain extreme conditions such as very high magnetic fields, $J$ is always negative. This does not change our argument since $J$ cannot change its sign.) We emphasize here that it is this unique constraint, $0\le J[\epsilon(t)]\le J_{\rm max}$ which renders the numerous compensating pulses developed in NMR literature inapplicable to this system. We also remark that although a pure $z$-axis rotation may be done by holding $h=0$ and $J$ constant, this is not desirable since one then loses access to universal control. This special case has been discussed in Ref.~\onlinecite{Wang.12}, and in the following we will assume $h>0$, and a composite pulse
is needed to perform $z$-axis rotations even without noise. For details, see Sec.~\ref{sec:zrot}. ($h$ and $J$ need not have the same sign. We assume $h>0$ only for convenience; our method applies equally well to the case of $h<0$. The only important thing is $h$ has to be a non-zero constant and $J$ has a definite sign.)

Rotations around both $x$- and $z$-axes are subject to decoherence. On one hand, fluctuations in the Overhauser field, for example the spin flip-flop induced by hyperfine interactions,  add a small, but unknown error term $\delta h$ to the Hamiltonian: $h\rightarrow h+\delta h$. On the other hand, the charge noise, caused by electrons hopping on and off impurity sites near the quantum dots, leads to deformation of the confinement potential and in turn the energy level structure.  As a consequence, errors will be introduced on the energy splitting between the singlet and triplet states, which we label by $\delta J$. This effect can alternatively be referred to as the control noise.

To treat these errors, we make a few assumptions. First, we assume that the control noise $\delta J$ and the magnetic field fluctuations $\delta h$ are uncorrelated, namely they are two independent sources of error. Second, we assume that $\delta J$ is completely caused by the fluctuations in the detuning, $\delta \epsilon$. Therefore,
\begin{equation}
\delta J\left[\epsilon\left(t\right)\right] = \delta \epsilon \left. \frac{\partial J\left(\epsilon\right)}{\partial \epsilon}\right|_{\epsilon = \epsilon\left(t\right)}=g(J)\delta\epsilon,\label{eq:dJdef}
\end{equation}
where $g(J)$ is a shorthand notation for $\partial J\left(\epsilon\right)/{\partial \epsilon}$ evaluated at the detuning that produces exchange $J$.
Third, the strong non-Markovian feature of the noises $\delta h$ and $\delta\epsilon$ allows us to assume that they are constant, albeit unknown, for the duration of a quantum gate. This last assumption is crucial since even dynamical decoupling would be impossible for completely white noise, and it is indeed the long time scale over which the noise varies compared to the very fast quantum gate operation times that allows us to perform corrected rotations. For a discussion of how our method works in the scenario where this third assumption is lifted, see Sec.~\ref{sec:benchmarking}.

Although the exact dependencies of $J$ on the detuning $\epsilon$ vary from sample to sample, an experimental fit gives the phenomenological dependence $J=J_1\exp(\epsilon/\epsilon_0)$, implying $g(J)\propto J$.\cite{Shulman.12,Dial.13} To facilitate our theoretical treatment, we will assume this form for $J(\epsilon)$ for most of the results given in this work. However, in Sec.~\ref{sec:altgofJ} we explicitly demonstrate that our method can easily accommodate other forms for $g(J)$.

We further assume that the pulses are square ``boxcar'' pulses with zero rise time, again for simplicity. However, in Sec.~\ref{sec:finiterisetime}, we show that our method continues to work well even in the case of finite rise times.

\section{Single-qubit operations}\label{sec:oneqrot}

\subsection{One-piece rotation: Rotation around axis $\boldsymbol{h\hat{x}+J\hat{z}}$}\label{sec:onepiece}

As discussed above, an (uncorrected) rotation by angle $\phi$ around the $x$-axis can be achieved by holding $J(t)$ at zero for a time $\phi/h$. In fact, holding $J(t)$ at a constant value $J(t)\equiv J$ would produce a rotation around the axis $h\hat{x}+J\hat{z}$. In the presence of both noise sources, such a rotation, which we denote by $U(J,\phi)$, has the form
\begin{align}
&U\left(J,\phi\right) \equiv \exp{\left[-i\left(\frac{h+\delta h}{2} \sigma_x + \frac{J+\delta J}{2} \sigma_z\right)\frac{\phi}{\sqrt{h^2+J^2}}\right]}\notag
\\
&= \exp{\left[-i\left(\frac{h}{2} \sigma_x + \frac{J}{2} \sigma_z\right)\frac{\phi}{\sqrt{h^2+J^2}}\right]}\left(I - i\sum_k\Delta_k \sigma_k \right)\notag\\
&\equiv R(J,\phi)\left(I - i\sum_k\Delta_k \sigma_k \right),\label{eq:UJphi}
\end{align}
where the sum on $k$ runs through $x,y,z$, and $R(J,\phi)$ is the desired (noiseless) operation. For convenience, we also define the ideal rotation by angle $\phi$ around an axis defined by the vector $\boldsymbol{r}$ as
\begin{equation}
R(\boldsymbol{r},\phi)=\exp\left(-i\frac{\boldsymbol{\sigma}\cdot\boldsymbol{r}}{|\boldsymbol{r}|}\frac{\phi}{2}\right).
\end{equation}
so that $R(J,\phi)=R(h\hat{x}+J\hat{z},\phi)$. Although this is perhaps a slight abuse of notation, it will prove very convenient.

To first order in $\delta h$ and $\delta J$, the error terms $\Delta_k$ are\cite{Kestner.13}
\begin{subequations}
\begin{align}
  \Delta_x &= \delta h \frac{h^2\phi + J^2\sin{\phi}}{2\left(h^2+J^2\right)^{3/2}} + \delta J \frac{h J\left(\phi-\sin{\phi}\right)}{2\left(h^2+J^2\right)^{3/2}} \\
  \Delta_y &= \delta h \frac{J\left(\cos{\phi}-1\right)}{2\left(h^2+J^2\right)} + \delta J \frac{h\left(1-\cos{\phi}\right)}{2\left(h^2+J^2\right)} \\
  \Delta_z &= \delta h \frac{h J\left(\phi-\sin{\phi}\right)}{2\left(h^2+J^2\right)^{3/2}} + \delta J \frac{\left(J^2\phi + h^2\sin{\phi}\right)}{2\left(h^2+J^2\right)^{3/2}}
\end{align}
\end{subequations}
Since $h$ is assumed to be held constant for the entire computation, we take $h=1$ as our energy unit for the remainder of the paper.

Our aim is to design a series of these pulses in such a way that the sum of all the error terms from each pulse equals zero. Our strategy, as shown in our previous works on {\sc supcode},\cite{Wang.12, Kestner.13} is to supplement the ``na\"ive'', uncorrected pulse of Eq.~\eqref{eq:UJphi} with a carefully chosen (uncorrected) identity operation $\widetilde{I}$, which has error
\begin{align}
\widetilde{I}^{(n)}= I - i \sum_k \delta_k \sigma_k\label{eq:ident},
\end{align}
such that the composite pulse $U\widetilde{I}$ is immune to the leading-order noise, namely, $\Delta_k+\delta_k=0$ for $k=x,y,z$ up to first order in $\delta h$ and $\delta J$ [the meaning of the superscript $(n)$ will become clear later]. To design such an identity, one typically needs to figure out the $\delta_k$ values corresponding to a given sequence and solve the coupled algebraic equations, $\Delta_k+\delta_k=0$ (which are typically nonlinear), to get the parameters that define the pulse sequence. This means that we need identities that contain a sufficient number of parameters such that there exist solutions to these equations.

There are infinitely many ways to perform an identity operation, but what we found most convenient is the interrupted $2\pi$ rotations \cite{Wang.12}, comprised of a $2\pi$ rotation interrupted by a $2\pi$ rotation about a different axis. In Ref.~\onlinecite{Kestner.13}, we present such an identity as
\begin{multline}
\!\!\!\widetilde{I}^{(n)}=U\left(j_n,m_n\pi - \theta_n\right)...U\left(j_1,m_1\pi - \theta_1\right) U\left(j_0,2m_0\pi\right)
\\
\times U\left(j_1,m_1\pi + \theta_1\right)...U\left(j_n,m_n\pi + \theta_n\right).\label{eq:leveln}
\end{multline}
where $m_k$ are integers, and $m_k\pi\pm\theta_k$ and $j_k$ are non-negative real numbers as required by the experimental constraints.  We refer to Eq.~\eqref{eq:leveln} as a ``level-$n$'' identity. This sequence of pulses is essentially a $2m_n\pi$ rotation around axis $\hat{x}+j_n\hat{z}$, interrupted by a $2m_{n-1}\pi$ rotation around another axis determined by $j_{n-1}$, with $\theta_n$ indicating the location of the interruption. The latter is, in turn, interrupted again by a $2m_{n-2}\pi$ rotation. This construction has several advantages. On one hand, it is guaranteed that in the absence of noise, this sequence becomes an exact identity. On the other hand, it makes searching for physically meaningful solutions easier because it allows for infinitely many degrees of freedom. In practice, one should start with a simpler identity. In principle, six parameters are all one would need to satisfy the noise cancelation conditions. However, due to the non-linearity of the equations, it is not guaranteed that the solutions are real and non-negative as required. When solutions are not found, one simply adds a level to acquire more tunable parameters. Therefore, this construction offers sufficient freedom that noise cancelation is always possible for all cases we studied, as will be explicitly demonstrated below.

With these comments, we can now outline the general procedure for generating pulse sequences corrected up to first order in the noise:
\begin{enumerate}
\item Determine the rotation $R(\hat{r},\phi)$ to be implemented. The pulses discussed in this section require $\hat{r}$ to be proportional to $\hat{x}+J\hat{z}$, i.e. ${\hat{r}=(\hat{x}+J\hat{z})/\sqrt{1+J^2}}$, for a certain physically allowed value of $J$.
\item Find the na\"ive pulse, $U(J,\phi)$ [Eq.~\eqref{eq:UJphi}] and its first order error terms $\Delta_{x,y,z}=\Delta_{x,y,z}^h\delta h+\Delta_{x,y,z}^J\delta J$.
\item Start with an uncorrected identity, say a level-3 identity $\widetilde{I}^{(3)}$ which has at least six parameters ($j_0$ through $j_3$, $\theta_1$ through $\theta_3$). Fix extra parameters such as $m_0$ through $m_3$ and one of the $j$ and $\theta$ parameters, making the total number of unknowns six. Again find the first order error terms $\delta_{x,y,z}=\delta_{x,y,z}^h\delta h+\delta_{x,y,z}^J\delta J$. However, here the coefficients $\delta_{x,y,z}^h$ and $\delta_{x,y,z}^J$ must contain six unknown parameters to be determined at the next stage.
\item Solve the six coupled equations $\Delta_{x,y,z}^h+\delta_{x,y,z}^h=0$ and $\Delta_{x,y,z}^J+\delta_{x,y,z}^J=0$.
\item If one has the desired solution (namely, all $j_k$ and angles $m_k\pi\pm\theta_k$ are non-negative real numbers), then the procedure is finished. \item Otherwise, try altering the fixed parameters in step 3, or if one still cannot find a satisfactory solution, increase the level of the identity in step 3.
\end{enumerate}

\begin{figure}[t]
    \centering
    \includegraphics[width=8cm, angle=0]{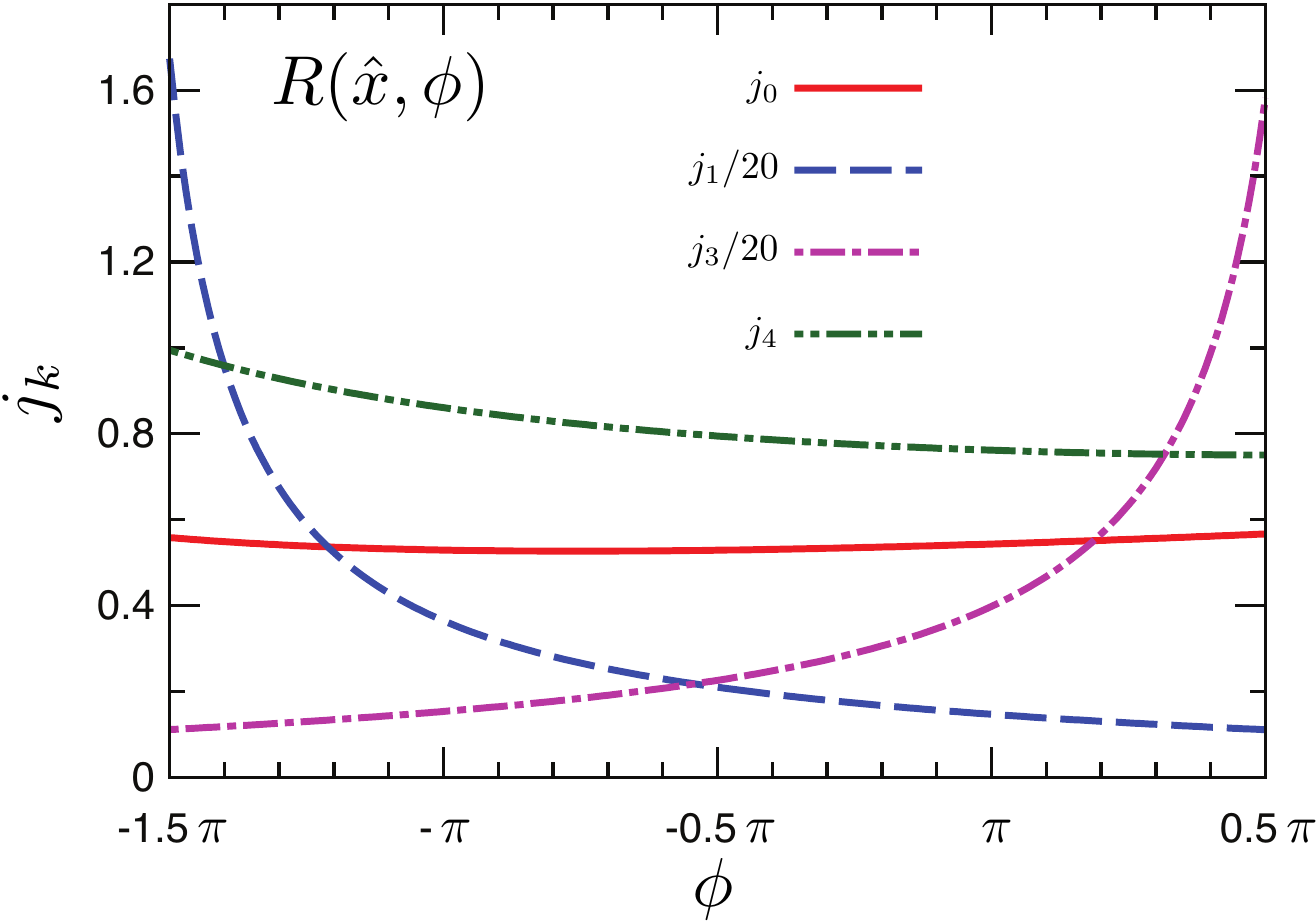}
    \caption{(Color online) Parameters for rotations around the $x$-axis for a range of angles, corresponding to the sequence shown in Eq.~\eqref{eq:OnePieceSeq}  with $j_2=J=0$. Note that $j_1$ and $j_3$ have been rescaled by a factor of 20 to fall into approximately the same range as the other parameters. In solving for these parameters, we have assumed  $g(J)=J/\epsilon_0$.}
    \label{fig:xpulse}
\end{figure}

There remain a few remarks to make. First, there are several ways to generate $\delta_{x,y,z}$ as functions of the pulse parameters. One may directly do a matrix multiplication, with all error terms analytically or numerically expressed for each trial solution to the equations. However a way we find most convenient in practice is to make use of the recursive nature of the identity design.\cite{Kestner.13} In other words, one may generate $\delta_k^{(n)}$ from $\delta_k^{(n-1)}$ ($k=x,y,z$), corresponding to level-$n$ and $n-1$ identities, while $\delta_k^{(0)}$ is known trivially. This typically eliminates the need for matrix multiplication, which leads to savings in computation time especially when the sequences become long.

Secondly, one does not always have to solve for six unknowns. By applying ``symmetric pulses'' from $t=0$ through $t=T_f$, i.e., $J(t)=J(T_f-t)$, one only needs to determine four unknown parameters since the coefficient of $\sigma_y$ is guaranteed to vanish,\cite{Wang.12} as shown below. This applies to rotations by any angle around an axis lying within the $x$-$z$ plane, which is obviously the case for what we study in this subsection. To see why the $\sigma_y$ component completely vanishes, consider that
\begin{align}
e^{-i(\sigma_x+J_k\sigma_z)t_k/2}\equiv A_k=a_{k0}{I}+a_{kx}{\sigma_x}+a_{kz}{\sigma_z}\label{eq:Ak}
\end{align}
Note that on the right hand side of Eq.~\eqref{eq:Ak}, $a_{k0}$, $a_{kx}$, $a_{kz}$ are arbitrary complex numbers, and there is no $\sigma_y$ term. Then, simple algebra reveals that given any operators $A_1$ and $A_2$ of the type of Eq.~\eqref{eq:Ak} with arbitrary coefficients, $A_2\cdot A_1\cdot A_2$ can also be written in such a form, free of $\sigma_y$ terms.
Applying this statement recursively to the time evolution for the entire sequence,  one immediately sees that for any $J(t)$ satisfying $J(t)=J(T_f-t)$, the resulting evolution operator $U$ does not contain a $\sigma_y$ component.

We are now ready to discuss how to correct a one-piece rotation, defined in Eq.~\eqref{eq:UJphi}, with {\sc supcode}. Let us reiterate our goal, that is to find an identity such that
\begin{align}
\widetilde{I}^{(5)}\cdot U\left(J,\phi\right)=e^{i\chi}R(\hat{x}+J\hat{z},\phi)
\left\{I+{\cal O}\left[\left(\delta h+\delta\epsilon\right)^2\right]\right\},
\label{eq:one-piececorr}
\end{align}
where $e^{i\chi}$ is an unimportant phase factor. We have found that a level-5 identity has enough degrees of freedom to cancel the noise. Of course, there are other identities which do the same job; we choose the particular identity shown in Eq.~\eqref{eq:one-piececorr} only for the sake of concreteness.

The next step is to find a way to make the entire pulse sequence ``symmetric''. This can be done as follows:
\begin{align}
&\hspace{0.5cm}\widetilde{I}^{(5)}\cdot U\left(J,\phi\right)\notag\\
&=U\left(J,\pi+\frac{\phi}{2}\right)\cdot\widetilde{I}^{(4)}\cdot U\left(J,\pi-\frac{\phi}{2}\right)U\left(J,\phi\right)\notag\\
&=U\left(J,\pi+\frac{\phi}{2}\right)\cdot\widetilde{I}^{(4)}\cdot U\left(J,\pi+\frac{\phi}{2}\right)\label{eq:one-piecesym}
\end{align}
Here if we make sure that $\theta_{1,2,3,4}=0$ then we have a symmetric pulse.
In this case, we can write the identity as
\begin{align}
\widetilde{I}^{(4)}=I-i\sigma_x(a_1\delta h+b_1\delta \epsilon)-i\sigma_z(a_3\delta h+b_3\delta \epsilon).\label{eq:identforonepiece}
\end{align}
where $a_{1,3}$ and $b_{1,3}$ are functions of the pulse parameters, namely they contain the information about how the identity is actually performed.

 \begin{figure}[t]
    \centering
    \includegraphics[width=8cm, angle=0]{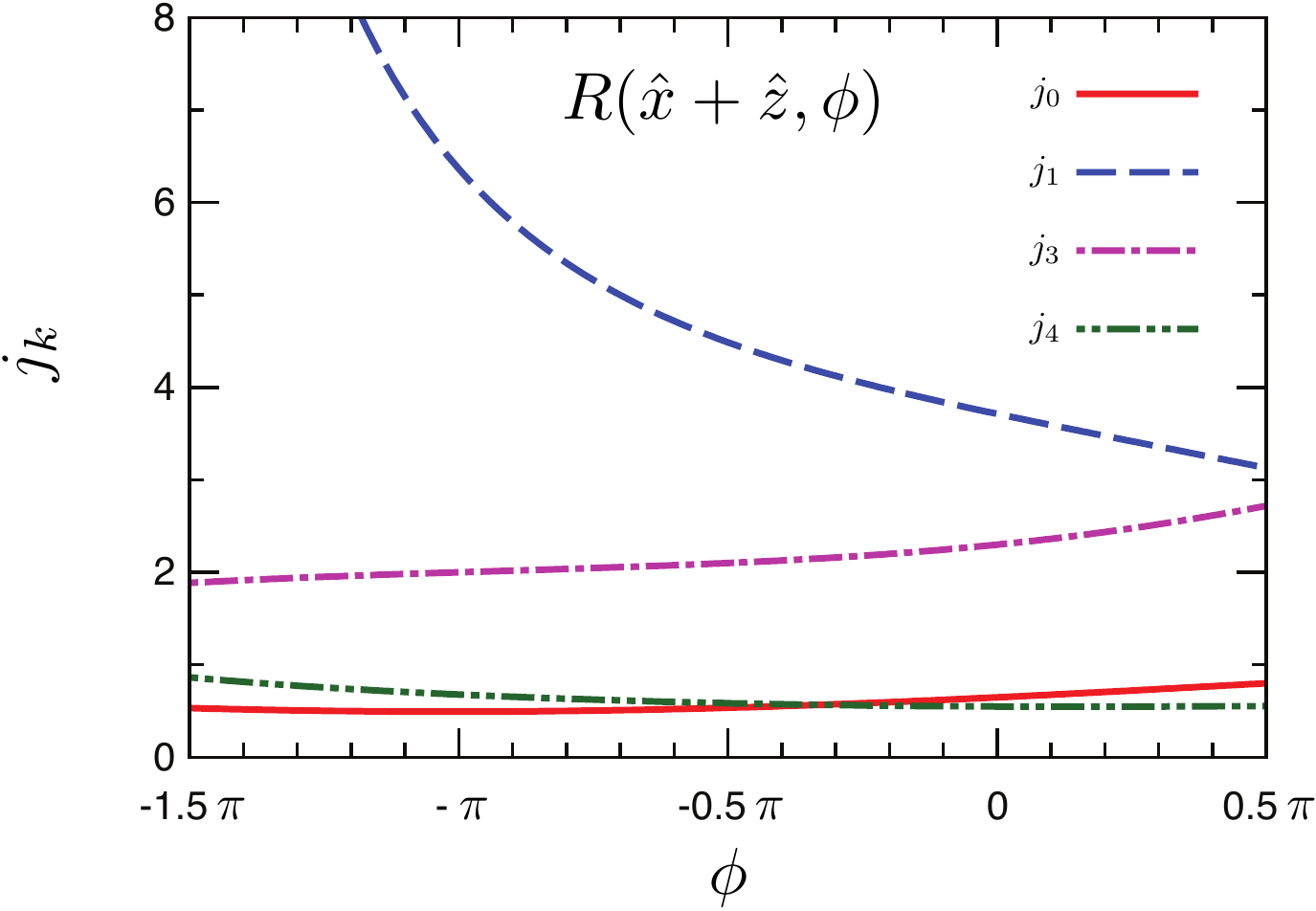}
    \caption{(Color online) Parameters for rotations around the axis $\hat{x}+\hat{z}$ for a range of angles, corresponding to the sequence shown in Eq.~\eqref{eq:OnePieceSeq}  with $j_2=0$, $J=1$. In solving for these parameters we have assumed  $g(J)=J/\epsilon_0$.}
    \label{fig:xpluszpulse}
\end{figure}

Plugging Eq.~\eqref{eq:identforonepiece} into Eq.~\eqref{eq:one-piecesym}, we have
\begin{align}
&U\left(J,\pi+\frac{\phi}{2}\right)\cdot\widetilde{I}\cdot U\left(J,\pi+\frac{\phi}{2}\right)\notag\\
=&-R(J,\phi)\Big\{I-i\big[(\alpha_1\delta h+\beta_1\delta\epsilon)\sigma_x+(\alpha_2\delta h+\beta_2\delta\epsilon)\sigma_y\notag\\
&\quad+(\alpha_3\delta h+\beta_3\delta\epsilon)\sigma_z\big]+{\cal O}\left[\left(\delta h+\delta\epsilon\right)^2\right]\Big\},
\end{align}
and direct algebra gives
\begin{widetext}
\begin{subequations}
\begin{align}
\displaybreak[1]
\alpha_1&=\frac{2 (a_1+a_3 J) \sqrt{1+J^2}+2 \pi +\phi +2 J (a_3-a_1 J) \sqrt{1+J^2} \cos\frac{\phi}{2}+J^2 \sin\phi}{2 \left(1+J^2\right)^{3/2}}\\
\beta_1&=\frac{2 \sqrt{1+J^2} \left[b_1+b_3 J+J (b_3-b_1 J) \cos\frac{\phi}{2}\right]+J g(J) (2 \pi +\phi -\sin\phi)}{2 \left(1+J^2\right)^{3/2}}\\
\alpha_2&=-\frac{\sin\frac{\phi}{2} \left[(a_3-a_1 J) \left(1+J^2\right)+J \sqrt{1+J^2} \sin\frac{\phi}{2}\right]}{\left(1+J^2\right)^{3/2}}\\
\beta_2&=-\frac{\sin\frac{\phi}{2} \left[(b_3-b_1 J) \left(1+J^2\right)-\sqrt{1+J^2} g(J) \sin\frac{\phi}{2}\right]}{\left(1+J^2\right)^{3/2}}\\
\alpha_3&=\frac{2 (a_1 J-a_3) \sqrt{1+J^2} \cos\frac{\phi}{2}+J \left[2 (a_1+a_3 J) \sqrt{1+J^2}+2 \pi +\phi -\sin\phi\right]}{2 \left(1+J^2\right)^{3/2}}\\
\beta_3&=\frac{2 \sqrt{1+J^2} \left[J (b_1+b_3 J)+(b_1 J-b_3) \cos\frac{\phi}{2}\right]+g(J) \left[J^2 (2 \pi +\phi )+\sin\phi\right]}{2 \left(1+J^2\right)^{3/2}}
\end{align}
\end{subequations}
\end{widetext}

Solving the coupled equations $\alpha_i=\beta_i=0$, ($i=1,2,3$) (note that only four out of six equations are independent), we have

\begin{subequations}
\begin{align}
\displaybreak[0]
a_1&=-\frac{2 \pi+\phi-2 J^2 \sin\frac{\phi}{2}}{2 \left(1+J^2\right)^{3/2}}\label{eq:onepieceeq1}
\\
b_1&=-\frac{J g(J) \left(2 \pi +\phi +2 \sin\frac{\phi}{2}\right)}{2 \left(1+J^2\right)^{3/2}}\\
a_3&=-\frac{J \left(2 \pi +\phi +2 \sin\frac{\phi}{2}\right)}{2 \left(1+J^2\right)^{3/2}}\\
b_3&=-\frac{g(J) \left[J^2 (2 \pi +\phi )-2 \sin\frac{\phi}{2}\right]}{2 \left(1+J^2\right)^{3/2}}\label{eq:onepieceeq4}
\end{align}
\end{subequations}

With these expressions it is then possible to find our composite pulse. We can construct the identity as
\begin{align}
\widetilde{I}^{(4)}&=U\left(j_4,\pi\right) U\left(j_3,\pi\right) U\left(j_2,\pi\right) U\left(j_1,\pi\right) U\left(j_0,4\pi\right)\notag\\
&\quad\times U\left(j_1,\pi\right) U\left(j_2,\pi\right) U\left(j_3,\pi\right) U\left(j_4,\pi\right).
\end{align}
(Note that we have already chosen $m_0=2$, $m_{1,2,3,4}=1$.) The entire pulse sequence reads
\begin{align}
&\widetilde{I}^{(5)}\cdot U\left(J,\phi\right)=\notag\\
&U\left(J,\pi+\frac{\phi}{2}\right)
U(j_4,\pi)
U(j_3,\pi)
U(j_2,\pi)
U(j_1,\pi)
U(j_0,4\pi)\notag\\
&\times U(j_1,\pi)
U(j_2,\pi)
U(j_3,\pi)
U(j_4,\pi)
U\left(J,\pi+\frac{\phi}{2}\right)
\label{eq:OnePieceSeq}
\end{align}

Since we only need to determine four parameters, and in Eq.~\eqref{eq:OnePieceSeq} there are five pulse parameters $j_0$ through $j_4$, the problem is under-constrained. Therefore, we may fix one pulse parameter, for example here we choose $j_2=0$. (Certainly other choices will work also, and we have explicitly verified that one can fix $j_2$ to a different value, or rather one may fix $j_1$ to be zero instead of $j_2$.) The parameters  $j_0$, $j_1$, $j_3$ and $j_4$ are found as follows.
For a desired rotation $R(J,\phi)$ with a known $g(J)$, one first finds $a_{1,3}$ and $b_{1,3}$ from
Eqs.~\eqref{eq:onepieceeq1}--\eqref{eq:onepieceeq4}.
Then, from the recursion relation one can find how $a_{1,3}$ and $b_{1,3}$ depend on the parameters $j_0$, $j_1$, $j_3$ and $j_4$, which are then solved for numerically from this set of equations. After that one verifies whether the solutions are physical, and if not the process is repeated with either a different assignment of the variables or other forms of the identity.

We show results of two representative cases: rotations around the $x$-axis, $R(\hat{x},\phi)$ (Fig.~\ref{fig:xpulse}), and rotations around axis $\hat{x}+\hat{z}$ (Fig.~\ref{fig:xpluszpulse}). In both figures we show solutions for a range of rotation angles covering a net rotation of $[0,2\pi)$ around that axis. In producing these results we have assumed $g(J)\propto J$, however as we emphasized above, our method works equally well for other forms of $g(J)$, as will be demonstrated in Sec.~\ref{sec:altgofJ}. For several important gates such as $R(\hat{x},-\pi/2)$, $R(\hat{x}, \pi)$, the identity operation $I=R(\hat{x}+\hat{z},2\pi)$, and the Hadamard gate $R(\hat{x}+\hat{z},\pi)$, we show explicitly the numerical values of the pulse parameters in Table~\ref
{tab:numericonepiece}. For $R(\hat{x},\pi/2)$, the numerical results presented in Fig.~\ref{fig:xpulse} require either $j_3\gtrsim30$ ($\phi=0.5\pi$) or $j_1\gtrsim30$ ($\phi=-1.5\pi$), which may be too large to access experimentally. However one can easily avoid this problem by using a slightly longer sequence:
\begin{align}
&U\left(J=0,\pi+\frac{\phi}{2}\right)
U(j_5,\pi)
U(j_4,\pi)
U(j_3,\pi)
U(j_2,\pi)\notag\\
&\times
U(j_1,\pi)U(j_0,4\pi)
U(j_1,\pi)
U(j_2,\pi)
U(j_3,\pi)
U(j_4,\pi)\notag\\
&\times
U(j_5,\pi)
U\left(J=0,\pi+\frac{\phi}{2}\right)
\label{eq:OnePieceSeqXRot}
\end{align}
with parameters shown in Table.~\ref{tab:numericxpion2}.
These gates form a subset of the Clifford gates, which are fundamental for quantum algorithms.

The results discussed in this section give corrected rotations around axes lying in a part of the first and third quadrant of the $x$-$z$ plane, bounded by the $x$-axis ($J=0$) and the axis $\hat{x}+J_{max}\hat{z}$. The larger the ratio $J/h$ is, the closer $\hat{x}+J\hat{z}$ comes to the $z$-axis. However since $J$ is also bounded from above, one cannot directly do a $z$-rotation with the results in this section. This will be discussed in the following section.

The duration of this pulse sequence, in terms of the total angle swept around the Bloch sphere, is roughly $12\pi\sim16\pi$. This is more than a factor of two shorter than our original {\sc supcode} sequence presented in Ref.~\onlinecite{Wang.12}. Moreover, within this much shorter time, we have achieved cancelation of both fluctuating Overhauser field gradients and charge noise simultaneously.

Finally, we remark again here that although it is not guaranteed that the nonlinear coupled equation array corresponding to a particular choice of parameters will have real and non-negative solutions, it is always possible to rearrange the parameters so that a physical solution may be found.

\begin{table}
 \centering
 \begin{tabular}{|c||c|c|c|c|c|c|c|}
 \hline
 & $J$ & $\phi$ & $j_0$ & $j_1$ & $j_2$ & $j_3$ & $j_4$\\
 \hline
 \hline
 $R(\hat{x},-\pi/{2})$ & 0 & $-\pi/{2}$ & 0.52870 & 4.1944 & 0 & 4.5149 & 0.79467 \\
 \hline
 $R(\hat{x},\pi)$ & 0 & $-\pi$ & 0.52902 & 7.2860 & 0 & 3.0639 & 0.86059 \\
 \hline
 \hline
 $I$ & 1 & 0 & 0.64714 & 3.7138 & 0 & 2.2988 & 0.54893\\
 \hline
 $R(\hat{x}+\hat{z},\pi)$ & 1 & $-\pi$ & 0.49263 & 6.3648 & 0 & 2.0008 & 0.67803 \\
 \hline
 \end{tabular}
 \caption{Parameters of the correcting sequence, Eq.~\eqref{eq:OnePieceSeq}, appropriate for several Clifford gates. Here the identity operation is achieved by $R(\hat{x}+\hat{z},2\pi)$. In solving for these parameters, we have assumed  $g(J)=J/\epsilon_0$.}\label{tab:numericonepiece}
\end{table}

\begin{table}
 \centering
 \begin{tabular}{|c||c|c|c|c|c|c|c|c|}
 \hline
 & $\phi$ & $j_0$ & $j_1$ & $j_2$ & $j_3$ & $j_4$ & $j_5$\\
 \hline
 \hline
 $R(\hat{x},\pi/2)$ & $\pi/{2}$ & 0.83930 & 0 & 1.1402 & 0.0025406 & 2.7063 & 0.46095\\
 \hline
 \end{tabular}
 \caption{Parameters of the correcting sequence Eq.~\eqref{eq:OnePieceSeqXRot}, appropriate for $R(\hat{x},\pi/2)$. In solving for these parameters, we have assumed  $g(J)=J/\epsilon_0$.}\label{tab:numericxpion2}
\end{table}

\subsection{$\boldsymbol{\hat{z}}$-axis rotation}\label{sec:zrot}

As discussed in the previous section, since we always have a non-zero $h$
(which is set to be the energy unit in this paper), one needs a composite pulse to achieve a $\hat{z}$-axis rotation even in the absence of noise. This is based on the following identity:\cite{HansonBurkard.07,Ramon.11}
\begin{align}
R(\hat{z},\phi)=-R(\hat{x}+\hat{z},\pi)R(\hat{x},\phi)R(\hat{x}+\hat{z},\pi).\label{eq:zrotzerothorder}
\end{align}
(We note here that this is not the only way of doing a $z$-axis rotation, and one may refer to Ref.~\onlinecite{Ramon.11} for more information).

Based on the results of Sec.~\ref{sec:onepiece}, we already have a composite pulse that cancels the noise from Eq.~\eqref{eq:zrotzerothorder}. Namely, we may correct each of the three terms on the right hand side of Eq.~\eqref{eq:zrotzerothorder} using the results of Sec.~\ref{sec:onepiece}. However, the resulting pulse sequence is long (around $40\pi\sim50\pi$ sweeps around the Bloch sphere). In this section, instead of correcting each of the three pieces, we shall try to do a ``one-shot'' correction, that is, correcting the right hand side of Eq.~\eqref{eq:zrotzerothorder} with only one identity. We have found that a level-6 identity is sufficient for our purpose. Compared to the length of three level-5 identities, a pulse sequence with only one level-6 identity is much shorter.

We first observe that a rotation around the $z$-axis also does not have $\sigma_y$ terms. Therefore, performing a symmetric pulse would reduce the number of equations to solve to four.
In order to utilize this nice feature, we insert the identity between $R(\hat{x},\phi)$ and $R(\hat{x}+\hat{z},\pi)$, but not at the right end of Eq.~\eqref{eq:zrotzerothorder}, so the corrected pulse looks like
\begin{align}
U(J=1,\pi)U(J=0,\phi)\widetilde{I}^{(6)}U(J=1,\pi).
\end{align}
As in Sec.~\ref{sec:onepiece}, the outer most level of the level-6 identity is absorbed into $U(J=0,\phi)$, so that the corrected pulse is
\begin{align}
U(J=1,\pi)U\left(J=0,\pi+\frac{\phi}{2}\right)\widetilde{I}^{(5)}\notag\\
\times U\left(J=0,\pi+\frac{\phi}{2}\right)U(J=1,\pi).\label{eq:zrotexpand}
\end{align}
We note here that an uncorrected identity operation can be placed anywhere, and here we have chosen a location which is most convenient, but other choices would also be possible.

Here, $\widetilde{I}^{(5)}$ has the same form as the right hand side of Eq.~\eqref{eq:identforonepiece}. We will not explicitly expand Eq.~\eqref{eq:zrotexpand},  but we note that our parameters in the identity operators are fixed by the rotation angle $\phi$ according to
\begin{subequations}
\begin{align}
a_1&=-\frac{1}{4}\left(4\pi+\sqrt{2}\pi+2\phi\right)\\
b_1&=-\frac{\pi}{2 \sqrt{2}}g(1)\\
a_3&=\frac{1}{8} \sec\frac{\phi}{2} \left(\sqrt{2}\pi+\sqrt{2}\pi\cos\phi+4\sin\phi\right)\\
b_3&=\frac{1}{8} \sec\frac{\phi}{2} \Big[g(1)\left(\sqrt{2}\pi+\sqrt{2}\pi\cos\phi-4\sin\phi\right)\notag\\
&\quad+4g(0)\cos\phi\Big]
\end{align}
\end{subequations}

\begin{figure}[]
    \centering
    \includegraphics[width=8cm, angle=0]{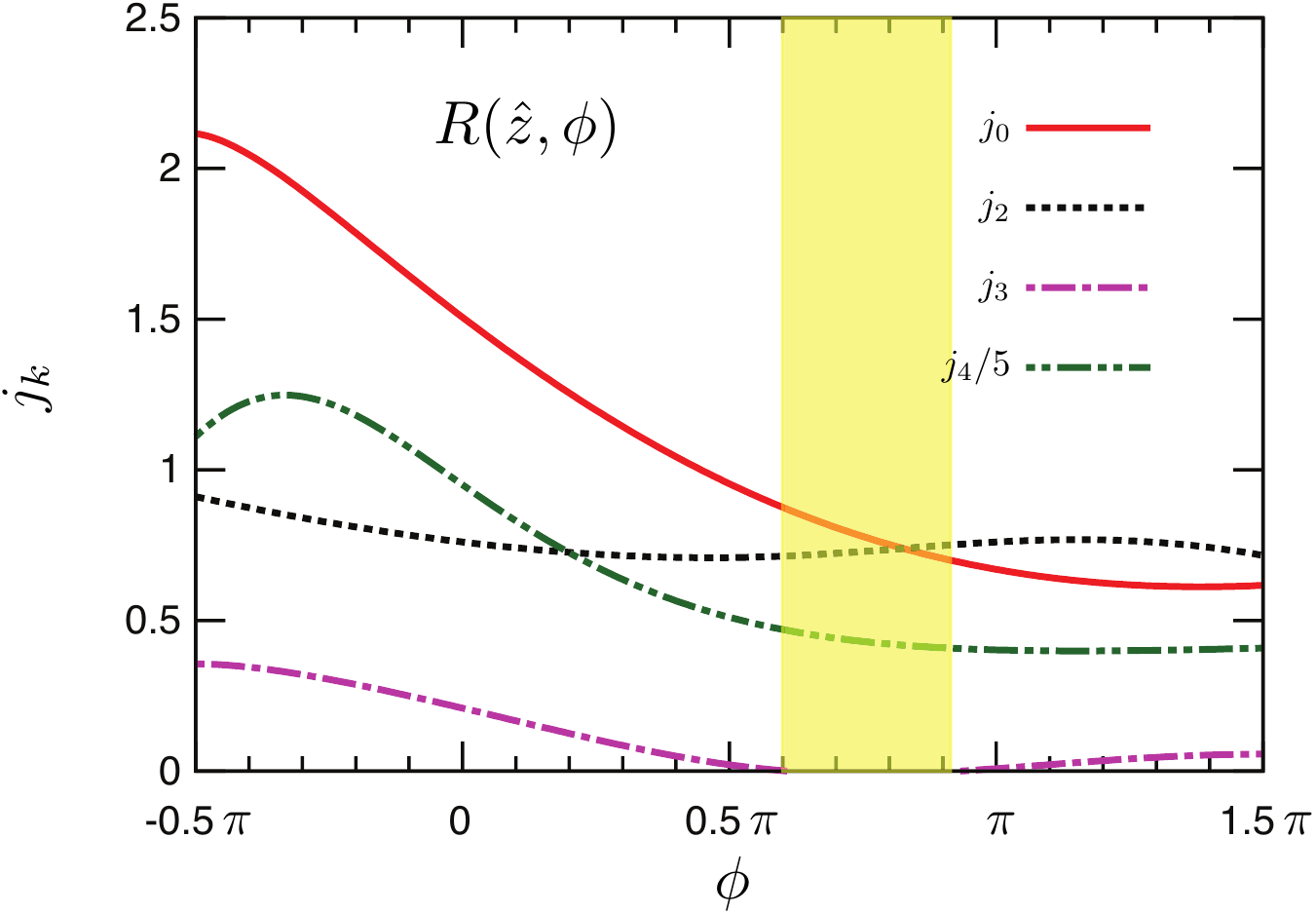}
    \caption{(Color online) Parameters for rotations around the $z$-axis for a range of angles, corresponding to the sequence shown in Eq.~\eqref{eq:zpulsecorr}. The yellow shaded area around $0.6\pi\lesssim\phi\lesssim0.9\pi$ indicates a range of $\phi$ for which the solutions of $j_3$ becomes negative and thus unphysical. In solving for these parameters we have assumed  $g(J)=J/\epsilon_0$.}
    \label{fig:zpulse}
\end{figure}

We consider the following sequence:
\begin{align}
\begin{split}
&U(J=1,\pi)
U(j_5=0,2\pi+\frac{\phi}{2})
U(j_4,\pi)
U(j_3,\pi)
U(j_2,\pi)\\
&\times
U(j_1=0,\pi)
U(j_0,4\pi)
U(j_1=0,\pi)
U(j_2,\pi)
U(j_3,\pi)\\
&\times
U(j_4,\pi)
U(j_5=0,2\pi+\frac{\phi}{2})
U(J=1,\pi)
\end{split}\label{eq:zpulsecorr}
\end{align}
(where we already inserted $\widetilde{I}^{(5)}$ into Eq.~\eqref{eq:zrotexpand}). Again, if we keep all of $j_0$ through $j_5$, the problem is under-constrained. Therefore we set $j_1=j_5=0$. The parameters $j_{0,2,3,4}$ are given in Fig.~\ref{fig:zpulse}. There is a shaded area between $0.6\pi\sim0.9\pi$ in Fig.~\ref{fig:zpulse}, which corresponds to unphysical solutions where $j_3$ is negative. To perform a rotation in this range, one may do two composite pulses that combine to give the desired rotation (namely to do a $0.8\pi$ net rotation by two back-to-back $0.4\pi$ net rotations). However this doubles the time duration of the sequence. An alternative is to re-assign the pulse parameters, for example set $j_2=j_4=0$ (instead of $j_1=j_5=0$ here). We have verified that this covers the range where physical solutions are missing in Fig.~\ref{fig:zpulse}. We also show numerical pulse parameters for Clifford gates in Table.~\ref{tab:zpulsepara}.

The pulse sequence of Eq.~\eqref{eq:zpulsecorr} requires about $18\pi\sim20\pi$ of rotation on the Bloch sphere, and is more than a factor of two shorter than the sequence that corresponds to correcting each of the three pieces of Eq.~\eqref{eq:zrotzerothorder} separately.

\begin{table}
 \centering
 \begin{tabular}{|c||c|c|c|c|c|c|c|}
 \hline
  & $\phi$ & $j_0$ & $j_1$ & $j_2$ & $j_3$ & $j_4$\\
 \hline
 \hline
 $R(\hat{z},-\pi/{2})$ & $-\pi/{2}$ & 2.1165 & 0 & 0.91080 & 0.35565 & 5.5498 \\
 \hline
 $R(\hat{z},\pi/{2})$ & $\pi/{2}$ & 0.95366 & 0 & 0.70853 & 0.021024 & 2.5518\\
 \hline
 $R(\hat{z},\pi)$ & $\pi$ & 0.66942 & 0 & 0.76034 & 0.0079157 & 2.0111\\
 \hline
 \end{tabular}
 \caption{
 Parameters of the correcting sequence, Eq.~\eqref{eq:zpulsecorr}, appropriate for Clifford $z$-rotations. In solving for these parameters, we have assumed  $g(J)=J/\epsilon_0$.
}\label{tab:zpulsepara}
\end{table}

\subsection{Arbitrary rotation}

Universal quantum computation requires complete single-qubit control, that is, the ability to perform arbitrary rotations around the Bloch sphere.
Such an arbitrary SU(2) rotation can be expressed as
\begin{equation}
\begin{pmatrix}
e^{i\alpha}\cos\theta & ie^{i\beta}\sin\theta \\
ie^{-i\beta}\sin\theta & e^{-i\alpha}\cos\theta
\end{pmatrix}
\end{equation}
and it is well known that it can be decomposed into an $z$-$x$-$z$ rotation\cite{NielsenChuang.00}
\begin{align}
R(\hat{z},\phi_a)R(\hat{x},\phi_b)R(\hat{z},\phi_c)
\end{align}
where the ``auxiliary angles'' are $\phi_a=\alpha+\beta$, $\phi_b=2\theta$, and $\phi_c=\alpha-\beta$.

It is straightforward to implement such a rotation since we already have $x$- and $z$-rotations. However the first step of optimization is made by noticing that a $z$-rotation is typically longer than an $x$-rotation, so we would like to rotate the entire coordinate frame around the $y$-axis by $\pi/2$, interchanging $x$ and $z$ axes. We note that the rotated general rotation
\begin{align}
\begin{split}
&R\left(\hat{y},\frac{\pi}{2}\right)\cdot
\begin{pmatrix}
e^{i\alpha}\cos\theta & ie^{i\beta}\sin\theta \\
ie^{-i\beta}\sin\theta & e^{-i\alpha}\cos\theta
\end{pmatrix}
\cdot R\left(\hat{y},-\frac{\pi}{2}\right)\\
=&
\begin{pmatrix}
 \cos\alpha\cos\theta+i \cos\beta\sin\theta& -i \cos\theta\sin\alpha-\sin\beta\sin\theta\\
 -i \cos\theta\sin\alpha+\sin\beta\sin\theta& \cos\alpha\cos\theta-i \cos\beta\sin\theta
\end{pmatrix}
\end{split}
\end{align}
remains an arbitrary rotation, which can be decomposed as
\begin{align}
R(\hat{x},\phi_a)R(\hat{z},\phi_b)R(\hat{x},\phi_c)
\end{align}
with auxiliary angles $\phi_a=-(\alpha+\beta)$, $\phi_b=2\theta$, and $\phi_c=\beta-\alpha$. According to Eq.~\eqref{eq:zrotzerothorder}, this decomposition can be written as
\begin{align}
R(\hat{x},\phi_a)R(\hat{x}+\hat{z},\pi)R(\hat{x},\phi_b)R(\hat{x}+\hat{z},\pi)R(\hat{x},\phi_c)\label{eq:arbseqzeroth}
\end{align}
up to a phase factor. Just as we discussed in Sec.~\ref{sec:zrot}, although we can correct each term on the right hand side of Eq.~\eqref{eq:arbseqzeroth} individually, we would prefer a ``one-shot'' correction at the cost of introducing a slightly longer uncorrected identity operation.

For an arbitrary rotation, $\sigma_y$ terms are in general present.  We therefore write the uncorrected identity operator as
\begin{align}
\begin{split}
\widetilde{I}&=I-i\sigma_x(a_1\delta h+b_1\delta \epsilon)-i\sigma_y(a_2\delta h+b_2\delta \epsilon)\\&\quad-i\sigma_z(a_3\delta h+b_3\delta \epsilon),
\end{split}\end{align}
and we insert the identity as
\begin{align}
\begin{split}
&U(J=0,\phi_a)U(J=1,\pi)\widetilde{I}U(J=0,\phi_b)\\
&\quad\times U(J=1,\pi)U(J=0,\phi_c).
\end{split}\label{eq:arbcorrschematic}
\end{align}
We choose to insert $\widetilde{I}$ in front of $U(J=0,\phi_b)$ because when the rotation axis is on the $x$-$z$ plane we can still make the pulse symmetric to simplify the problem. Note that the rotations discussed in Secs.~\ref{sec:onepiece} and \ref{sec:zrot} do not cover the entire plane. For $h>0$ and $J\ge0$, only rotations with the axis in a region of the first quadrant can be corrected with the methods of Secs.~\ref{sec:onepiece}. Again, one can choose other forms of the sequence, and the infinitely many degrees of freedom of the nested identities ensures that finding physical solutions to the coupled equations is always possible.

We will not explicitly expand Eq.~\eqref{eq:arbcorrschematic}, as the result would be rather complicated. To give an outline for how one can generate a corrected rotation around some arbitrarily chosen axis, one first solves, to zeroth order, for the auxiliary angles $\phi_{a,b,c}$ corresponding to this rotation axis, and then picks a level-$n$ identity with a sufficient number of degrees of freedom. The errors resulting from the uncorrected identity are fixed by
\begin{subequations}
\begin{align}
\displaybreak[1]
a_1&=-\frac{1}{4}\left(\sqrt{2}\pi+2\phi_b\right)\\
b_1&=-\frac{1}{4}\left[\sqrt{2}\pi g(1)+2g(0)(\sin\phi_a+\sin\phi_c)\right]\\
a_2&=\frac{1}{8}\Big(4-4\cos\phi_b+\sqrt{2}\pi\sin\phi_b+4\phi_c\sin\phi_b\Big)\\
b_2&=\frac{1}{8}\Big\{4g(0)[2-\cos\phi_a+\cos\phi_b(\cos\phi_c-2)]\notag\\
&\quad+g(1)(4\cos\phi_b-4+\sqrt{2}\pi\sin\phi_b)\Big\}\\
a_3&=-\frac{1}{8}\Big[\sqrt{2}\pi(1+\cos\phi_b)-4(\phi_a+\phi_c\cos\phi_b+\sin\phi_b)\Big]\\
b_3&=-\frac{1}{8}\Big\{\sqrt{2}\pi g(1)(1+\cos\phi_b)\notag\\
&\quad-4\sin\phi_b\left[(\cos\phi_c-2)g(0)+g(1)\right]\Big\}
\end{align}
\end{subequations}

One then equates these error terms to the ones generated from the concrete form of the identity, which contains all the pulse parameters. Although it is certainly harder to find physical solutions for this six-equation system, our experience is that one is always able to find one since the degrees of freedom can always be increased. One useful remark is that additional degrees of freedom come from adding $2\pi$'s to the auxiliary angles, or from reversing the order of $\phi_{a,b,c}$ since $R(\hat{r},-\phi)=R(\hat{r},2\pi-\phi)$. In the end, there are many sets of solutions for $\phi_{a,b,c}$, and there are many parameters in the uncorrected identity operation, so one is almost always guaranteed to have enough degrees of freedom when the level of the identity is made high enough.

\begin{figure}[]
    \centering
    \includegraphics[width=8cm, angle=0]{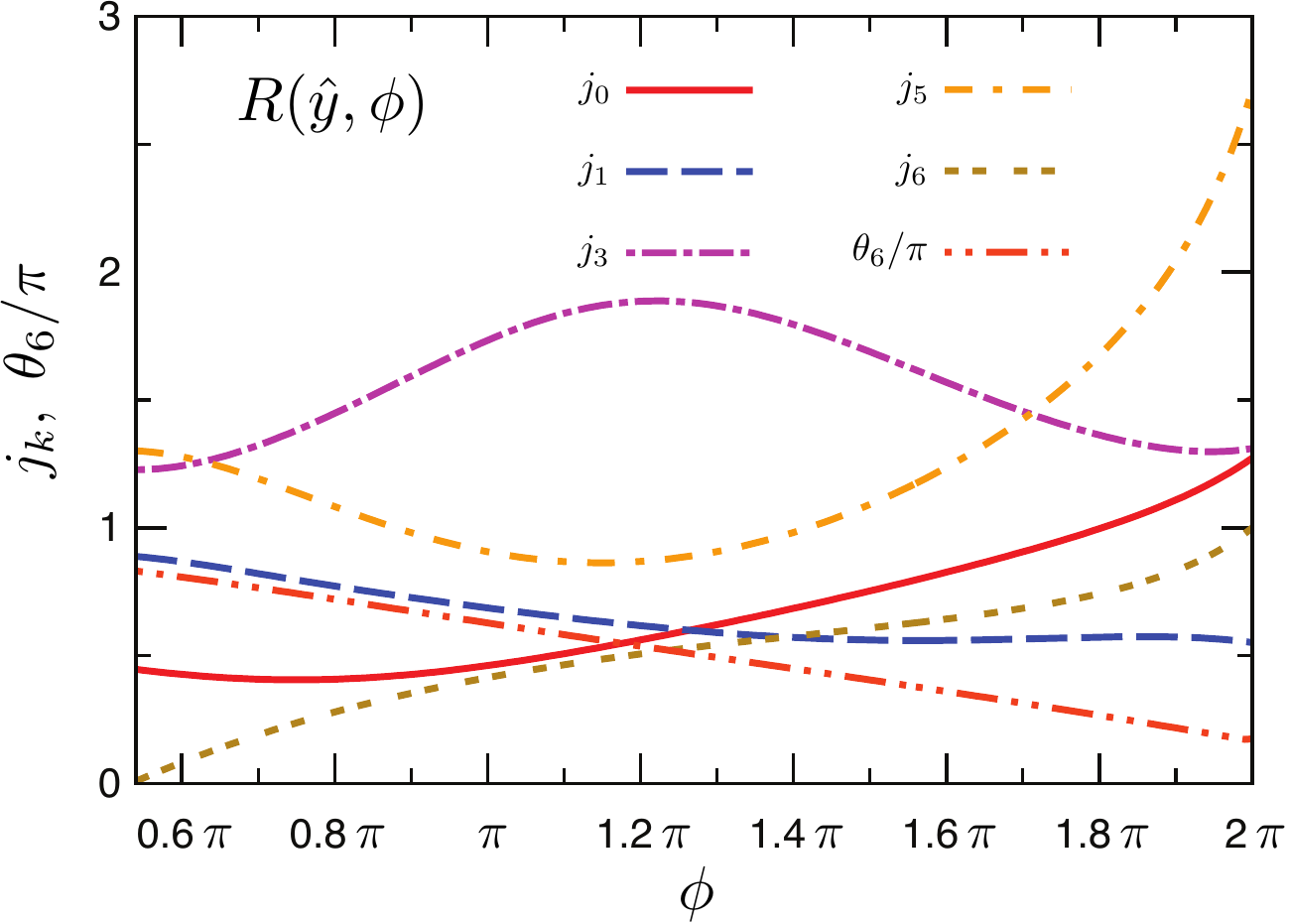}
    \caption{(Color online) Parameters for rotations around the $y$-axis for a range of angles, corresponding to the sequence shown in Eq.~\eqref{eq:ypulsecorr}. For $\phi\lesssim0.53\pi$, $j_6$ becomes negative. In solving for these parameters, we have assumed  $g(J)=J/\epsilon_0$.}
    \label{fig:ypulsepara}
\end{figure}

In Fig.~\ref{fig:ypulsepara} we show numeric values of parameters for an $R(\hat{y},\phi)$ rotation,  corresponding to the sequence
\begin{align}
\begin{split}
&U(J=0,\phi_a=\frac{3\pi}{2})
U(J=1,\pi)
U(j_6,\pi-\theta_6)
U(j_5,\pi)\\
&\times U(j_4,\pi)
U(j_3,\pi)
U(j_2=0,\pi)
U(j_1,\pi)
U(j_0,4\pi)U(j_1,\pi)\\
&\times
U(j_2=0,\pi)
U(j_3,\pi)
U(j_4,\pi)
U(j_5,\pi)
U(j_6,\pi+\theta_6)\\
&\times
U(J=0,\phi_b=\phi)
U(J=1,\pi)
U(J=0,\phi_c=\frac{\pi}{2}).
\end{split}\label{eq:ypulsecorr}
\end{align}
At least one of the six parameters to be solved for needs to be $\theta_n$, which breaks the ``time-reversal'' symmetry of the uncorrected identities. Here, we consider a level-6 identity with $\theta_6$ and $j_0$ through $j_6$ to be determined. (We take $j_2=0$ to keep the number of unknown variables six.)
This kind of rotation is an important operation used, for example, in converting the two-qubit Ising gate to a {\sc cnot} gate.\cite{Klinovaja.12,Li.12}
We remark that the pulse sequence of Eq.~\eqref{eq:ypulsecorr} sweeps a total angle of $20\pi\sim22\pi$ around the Bloch sphere. Compared to a na\"ive correction of an $x$-$z$-$x$ sequence, which would cost $40\pi\sim50\pi$, this is again a factor of two improvement.

\begin{table*}
 \centering
 \begin{tabular}{|c||c|c|c|c|c|c|c|c|c|c|c|}
 \hline
  & $j_0$ & $j_1$ & $j_2$ & $j_3$ & $j_4$ & $j_5$ & $j_6$ & $\theta_6$ & $\phi_a$ &  $\phi_b$ & $\phi_c$ \\
 \hline
 \hline
 $R(\hat{y},-\pi/{2})$ & 0.75330 & 0.56113 & 0 & 1.6884 & 0& 1.0914 & 0.60835 & 1.2726 & $3\pi/2$ & $3\pi/2$ & $\pi/2$ \\
 \hline
 $R(\hat{y},\pi/{2})$ & 0.81782 & 0 & 1.3113 & 0.55040 & 1.0366 & 0 & 1.6911 & -1.1929 & $5\pi/2$ & $3\pi/2$ & $3\pi/2$\\
 \hline
 $R(\hat{y},\pi)$ & 0.46134 & 0.68677 & 0 & 1.7332 & 0 & 0.90639 & 0.41421 & 1.9727 & $3\pi/2$ & $\pi$ & $\pi/2$\\
 \hline
 \hline
$R(\hat{x}-\hat{z},\pi)$ & 0.71967 & 1.3078 & 0 & 0.81623 & 0 & 1.5118 &  0 & $-3\pi/4$ & $\pi/2$ & $3\pi/2$ & $\pi/2$\\
\hline
$R(\hat{x}+\hat{y},\pi)$ & 0.54448 & 0.63330 & 0 & 1.4188 & 0 & 1.7652 & 0.041400 & 1.7384 & 0& $\pi/2$ & $3\pi$\\
\hline
$R(\hat{x}-\hat{y},\pi)$ & 0.60618 & 0.71995 & 0 & 0.88507 & 0 & 2.2037 & 0.019841 & 2.1125 & $\pi$ & $5\pi/2$ & $2\pi$\\
\hline
$R(\hat{y}+\hat{z},\pi)$ & 1.1424 & 0 & 0.59501 & 0.0042383 & 1.4268 & 0 & 0.62132 & 2.1010 & $7\pi/2$ & $\pi$ & $2\pi$\\
\hline
$R(\hat{y}-\hat{z},\pi)$ & 0.31843 & 0.84663 & 0 & 1.2694 & 0 & 0.92116 & 0.20711 & 1.7924 & $\pi/2$ & $\pi$ & 0\\
\hline
\hline
$R(\hat{x}+\hat{y}+\hat{z},2\pi/3)$ & 0.40554 & 1.1271 & 0 & 1.0423 & 0 & 1.1682 & 0.022417 & 2.0737 & 0 & $\pi/2$ & $\pi/2$ \\
\hline
$R(\hat{x}+\hat{y}+\hat{z},4\pi/3)$ & 1.1099 & 0.67185 & 0 & 0.58455 & 0 & 3.5271 &  0.72636 & 1.3825 & $7\pi/2$ & $7\pi/2$ & $2\pi$\\
\hline
$R(\hat{x}+\hat{y}-\hat{z},2\pi/3)$ & 0.81495 & 0 & 0.53383 & 0.16963 &
1.0824 & 0 & 0.73536 & -1.7509 & $5\pi/2$ & $3\pi/2$ & $4\pi$\\
\hline
$R(\hat{x}+\hat{y}-\hat{z},4\pi/3)$ & 0.46515 & 0.90353 & 0 & 1.2451 & 0 &
1.2943 & 0.035404 & 1.8526 & 0 & $\pi/2$ & $3\pi/2$\\
\hline
$R(\hat{x}-\hat{y}+\hat{z},2\pi/3)$ & 0.59703 & 0.74094 & 0 & 0.88895 & 0 &
2.0930 & 0.029762 & 1.9536 & $\pi/2$ & $5\pi/2$ & $2\pi$\\
\hline
$R(\hat{x}-\hat{y}+\hat{z},4\pi/3)$ & 0.96348 & 1.0402 & 0 & 0.47533 & 0 & 5.0237 & 0.19295 & -2.2348 & $2\pi$ & $7\pi/2$ & $3\pi/2$ \\
\hline
$R(-\hat{x}+\hat{y}+\hat{z},2\pi/3)$ & 0.52445 & 0.69563 & 0 & 1.36738 & 0 & 1.6155 & 0.0095420 & 2.2507 & $3\pi/2$ & $\pi/2$ & $2\pi$ \\
\hline
$R(-\hat{x}+\hat{y}+\hat{z},4\pi/3)$ & 1.3517 & 0.79872 & 0 & 0.40171 & 0 & 8.0500 & 0.97474 & 1.5893 & $4\pi$ & $7\pi/2$ & $5\pi/2$\\
\hline
 \end{tabular}
 \caption{
Parameters of the correcting sequence, Eq.~\eqref{eq:FivePieceSeq}, appropriate for all remaining Clifford gates not discussed in Sec.~\ref{sec:onepiece} and Sec.~\ref{sec:zrot}.}\label{tab:genpulsepara}
\end{table*}

A general rotation around any arbitrary axis can be achieved by the following sequence
\begin{align}
\begin{split}
&
U(0,\phi_a)
U(1,\pi)
U(j_6,\pi-\theta_6)
U(j_5,\pi)
U(j_4,\pi)
U(j_3,\pi)\\
&\times
U(j_2,\pi)
U(j_1,\pi)
U(j_0,4\pi)
U(j_1,\pi)
U(j_2,\pi)
U(j_3,\pi)\\
&\times
U(j_4,\pi)
U(j_5,\pi)
U(j_6,\pi+\theta_6)
U(0,\phi_b)
U(1,\pi)
U(0,\phi_c).
\end{split}\label{eq:FivePieceSeq}
\end{align}
We show numerical values of parameters for the Clifford gates in
Table~\ref{tab:genpulsepara}. In finding these pulses, we mostly fix $j_2=j_4=0$, but there are a few cases where we have fixed $j_1=j_5=0$ instead. For example, if we use $j_2=j_4=0$ in the search for the pulse that implements $R(\hat{y},\pi/2)$, the $j_6$ value becomes negative, as is clear from Fig.~\ref{fig:ypulsepara}; while if we fix $j_1=j_5=0$, then a set of physical solutions can be found. One may also notice that for $R(\hat{x}-\hat{z},\pi)$, $\theta_6=-\phi_b/2$, and there are only four remaining $j$ values to solve for. This is because the rotation axis is within the $x$-$z$ plane, and choosing this $\theta_6$ value makes the entire sequence symmetric.

\subsection{Application to general $J(\epsilon)$}\label{sec:altgofJ}

For most calculations shown in this work, we have assumed $J(\epsilon)=\exp(\epsilon/\epsilon_0)$, implying $g(J)=J/\epsilon_0$. While this is a widely used model, it is not necessarily applicable to any double quantum dot sample. For example, in this model, when the detuning is tuned far toward the negative side, the exchange interaction vanishes. However, there may be residual tunneling between the two dots, so that one may not completely turn off the exchange interaction, and the minimal value of $J$ would be a positive number $J_{\rm min}$. Also, the exponential dependence of $J$ on $\epsilon$ merely reflects the shape of the avoided crossing where, as one increases $\epsilon$, $J(\epsilon)$ increases much faster than linearly.  Therefore, in this subsection we demonstrate that our method would work for other forms $J(\epsilon)$. The functional form of $J(\epsilon)$ generally varies from sample to sample and must be measured individually before one optimizes the {\sc supcode} sequence, and it is not possible to present results for all cases. Here, we will present numerical results for two cases of $J(\epsilon)$ for one particular operation, $R(\hat{x}+\hat{z},\phi)$.

\begin{figure}[t]
    \centering
        \includegraphics[width=8cm, angle=0]{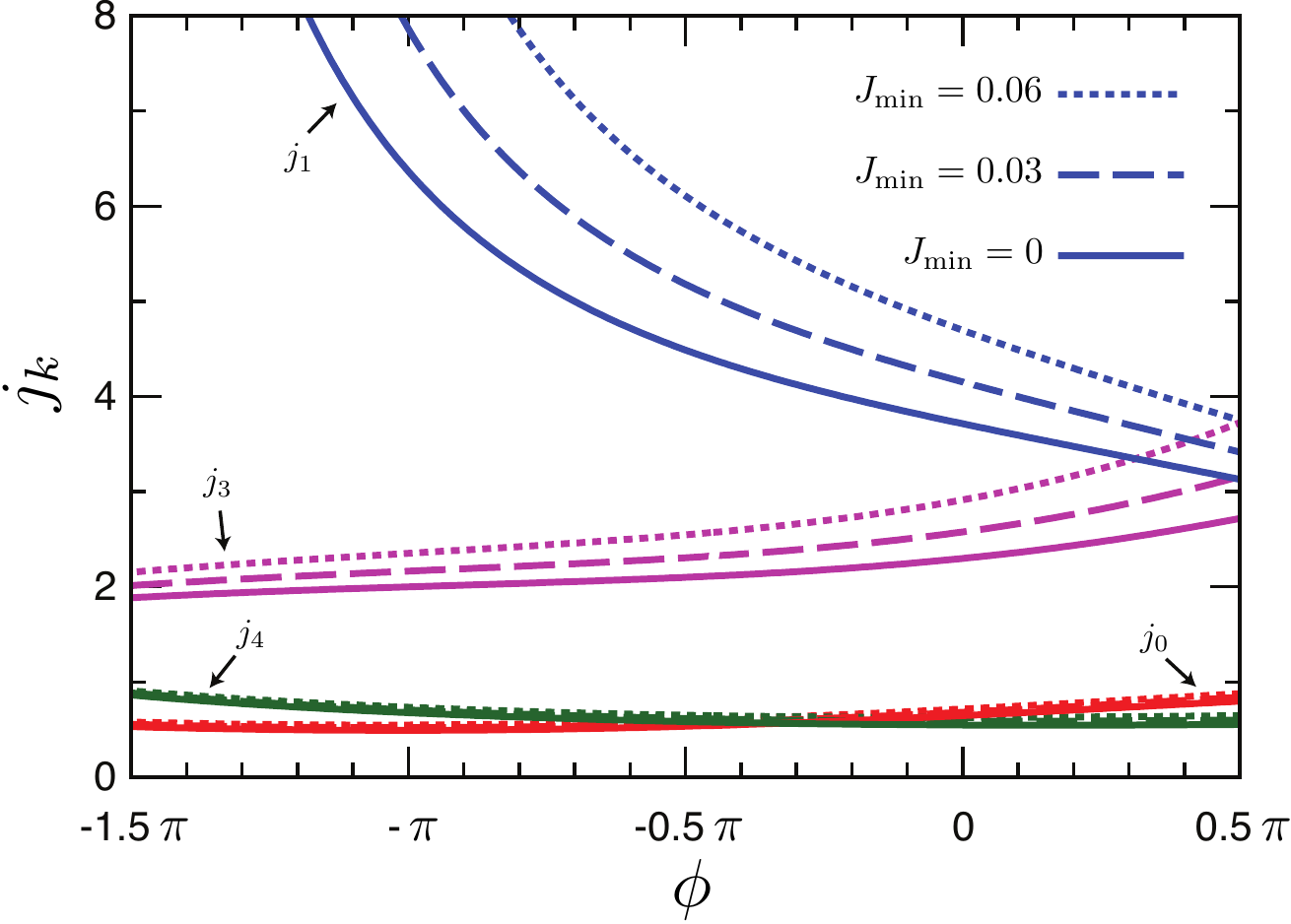}
    \caption{(Color online) Parameters for rotations around the axis $\hat{x}+\hat{z}$ v.s. rotation angle for several $J_{\rm min}$. In solving for these parameters, we have assumed  $g(J)=(J-J_{\rm min})/\epsilon_0$. The solid lines are for $J_{\rm min}=0$ and are the same as what is shown in Fig.~\ref{fig:xpluszpulse}. The dashed lines and dotted lines are for $J_{\rm min}=0.03$ and 0.06 respectively as indicated on the figure. As in Fig.~\ref{fig:xpluszpulse}, the parameters shown
  correspond to the sequence Eq.~\eqref{eq:OnePieceSeq}  with $J=1$, but $j_2=J_{\rm min}$.}
    \label{fig:altgofJJmin}
\end{figure}

We first consider the case of $J(\epsilon)=J_{\rm min}+\exp(\epsilon/\epsilon_0)$, corresponding to
\begin{align}
g(J)=(J-J_{\rm min})/\epsilon_0.
\end{align}
We would like to ask the question: how much will the parameters for the {\sc supcode} sequence change as one turns on $J_{\rm min}$? We show the results in Fig.~\ref{fig:altgofJJmin}. The solid lines are for $J_{\rm min}=0$, which are exactly the same data as what is shown in Fig.~\ref{fig:xpluszpulse}. The dashed lines and dotted lines are for $J_{\rm min}=0.03$ and 0.06 respectively. We see that when we have a positive $J_{\rm min}$, the parameters undergo only relatively small shifts away from their original values. This indicates that the solution we have found is also robust. In practice, it is also possible to take the ideal solution for $J_{\rm min}=0$ as the starting point, and then search for the optimal solution corresponding to the true $g(J)$ using some experimental measure of fidelity.

\begin{figure}[t]
    \centering
        \includegraphics[width=8cm, angle=0]{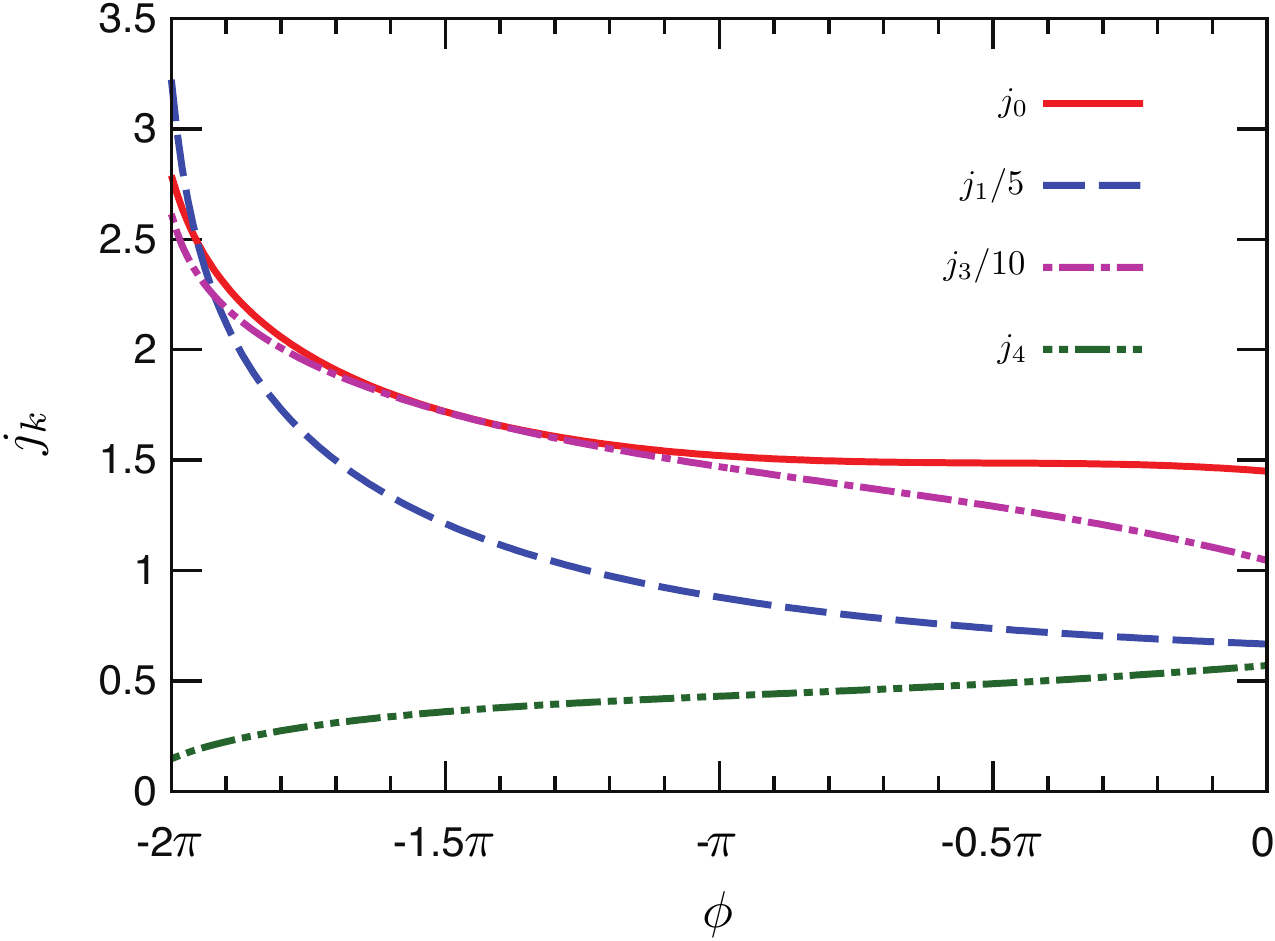}
    \caption{(Color online) Parameters for rotations around the axis $\hat{x}+\hat{z}$ v.s. rotation angle for $J(\epsilon)$ as defined in Eq.~\eqref{eq:altgJJarb}. Parameters in Eq.~\eqref{eq:altgJJarb} are $J_{\rm min}=0.008$, $J_1=67.3$, $\alpha_1=0.476$, $\alpha_2=0.156$, $\gamma=0.812$. $j_1$ and $j_3$ are rescaled by a certain factor as indicated in the figure. As in Figs.~\ref{fig:xpluszpulse} and \ref{fig:altgofJJmin}, the parameters shown
  correspond to the sequence Eq.~\eqref{eq:OnePieceSeq}  with $J=1$ and $j_2=0.01>J_{\rm min}$.}
    \label{fig:altgofJJarb}
\end{figure}

We next consider a somewhat arbitrarily chosen case with
\begin{align}
J(\epsilon)=J_{\rm min}+J_1e^{\displaystyle -(\epsilon/\alpha_1+\sqrt{\epsilon}/\alpha_2)^\gamma}.\label{eq:altgJJarb}
\end{align}
To solve for the pulse parameters, we first find $g(J)$, which has a complicated form we will not show explicitly, and then take the solution for the simple $g(J)\propto J$  case as the starting point for the solution search. Remarkably, this process always converges, and we show results for a representative case in Fig.~\ref{fig:altgofJJarb}. Although the line shape drastically changes, which is expected because we have a completely different $g(J)$, we emphasize that we still find a physical solution, and that this solution can easily be found after we take the known solution for the $g(J)\propto J$ case as the algorithm input. This means that our method should work seamlessly for alternative choices of $J(\epsilon)$, as long as the form of this function is known for the specific sample to be measured.

\subsection{Finite rise time}\label{sec:finiterisetime}

\begin{figure}[t]
    \centering
        \includegraphics[width=8cm, angle=0]{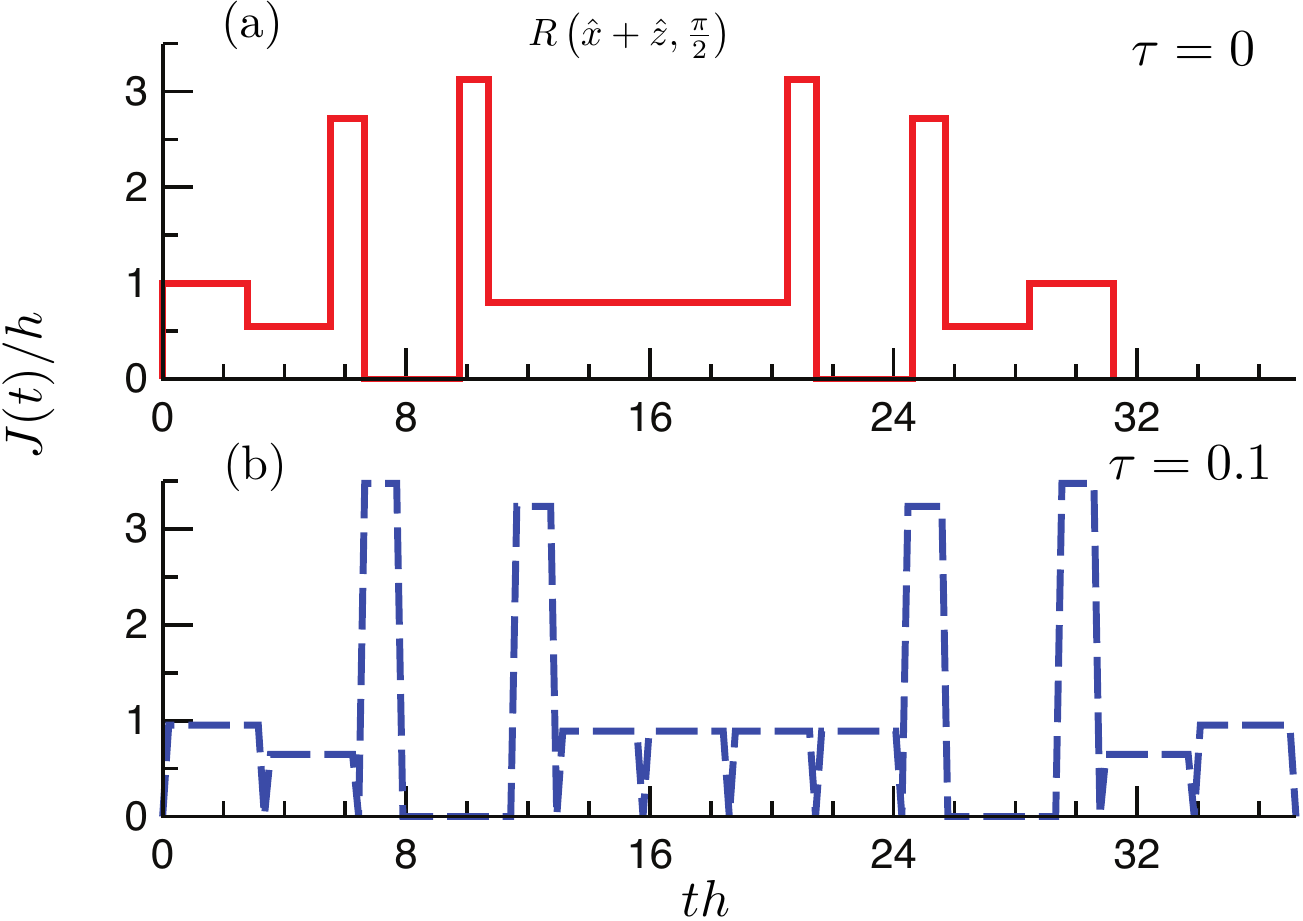}
    \caption{(Color online) Pulse shape for $R(\hat{x}+\hat{z},\pi/2)$ for (a) square pulse (no rise time, $\tau=0$) and (b) trapezoidal pulse with finite rise time $\tau=0.1$. In solving for these parameters, we have assumed  $g(J)=J/\epsilon_0$. Note that for panel (a), the pulse sequence ends at around $T_f=31.24$, and for panel (b) the sequence ends at $T_f=37.22$. }
    \label{fig:finiterizeshape}
\end{figure}

\begin{figure}[t]
    \centering
        \includegraphics[width=8cm, angle=0]{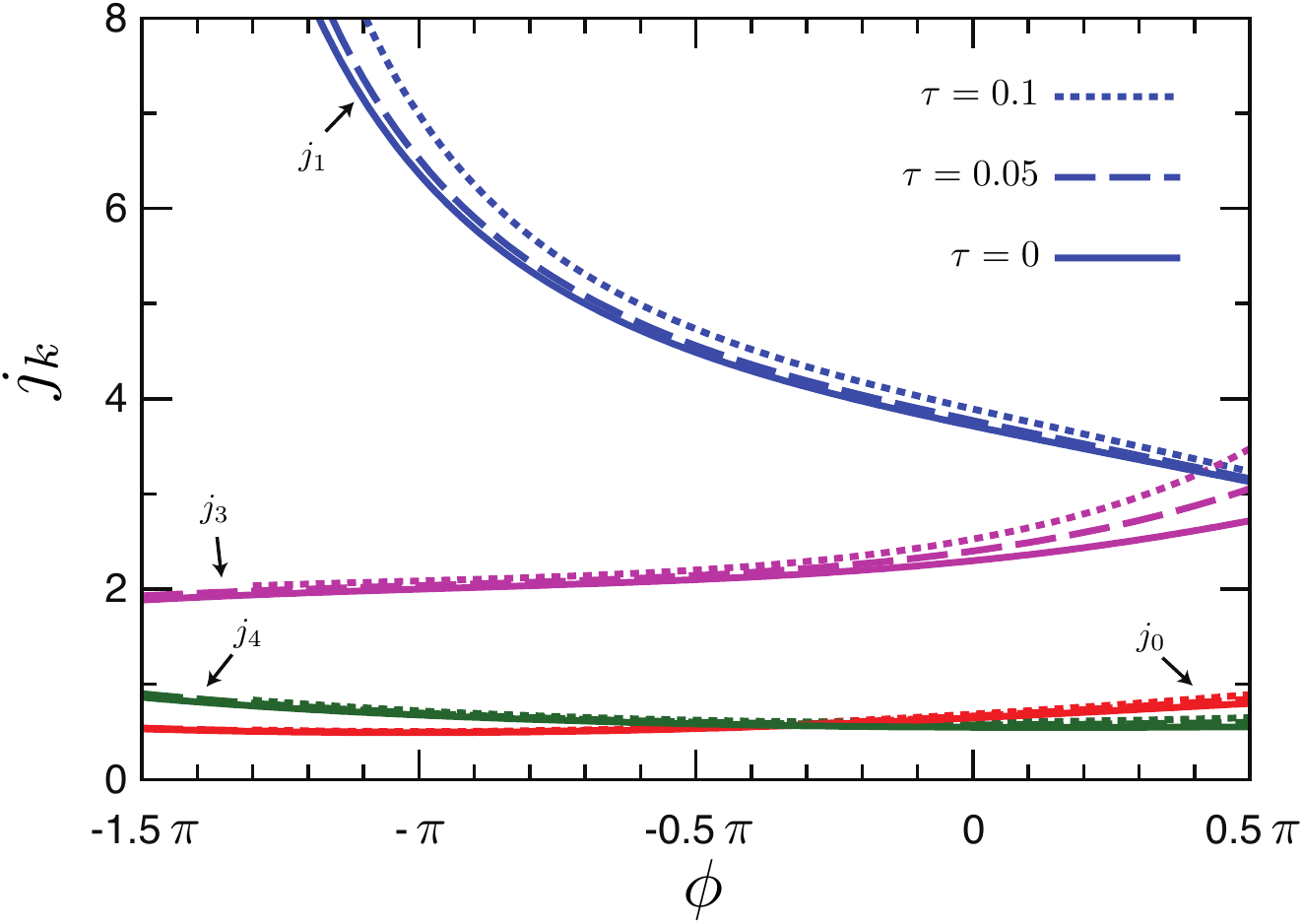}
    \caption{(Color online) Parameters for rotations around the axis $\hat{x}+\hat{z}$ for a range of angles, when a finite pulse rise time $\tau$ as defined in Eqs.~\eqref{eq:finiterise1} and \eqref{eq:finiterise2} is taken into consideration. The parameters correspond to the sequence shown in Eq.~\eqref{eq:OnePieceSeq} but with a finite pulse rise time $\tau$ incorporated in a similar manner as  Fig.~\ref{fig:finiterizeshape}(b). Other parameters not shown are $j_2=0$ and $J=1$. In solving for these parameters, we have assumed  $g(J)=J/\epsilon_0$.}
    \label{fig:finiterizepara}
\end{figure}

Thus far, for the convenience of the theoretical treatment, we have assumed that the pulses have rectangular shapes, i.e., $J$ can be turned on and off instantaneously. In Ref.~\onlinecite{Wang.12}, we discussed how the parameters of the pulses presented in that work would change if one takes into consideration the finite rise times of the pulses, and the conclusion was that the parameters would change very slightly so that one may simply perform a local search around the solution for the ideal, zero rise time, case. In this subsection, we demonstrate that this is also the case for the pulse sequences discussed here and in Ref.~\onlinecite{Kestner.13}, which are capable of correcting both $\delta h$ and $\delta J$ error.

We introduce a function $f(t,T)$ which encapsulates the information about the pulse shape. The rotation around the axis $\hat{x}+J\hat{z}$ is then expressed as
\begin{equation}
\check{R}( \hat{x} + J\hat{z};\phi)=\mathcal{T}\exp\biggl[-i \int_0^T \Bigl(\frac{1}{2} \sigma_x + \frac{J\cdot f(t,T)}{2} \sigma_z\Bigr)\,\mathrm{d}t\biggr]\quad,\label{eq:finiterise1}
\end{equation}
where $\mathcal{T}$ is the time-ordering operator.

As an example, we consider pulses with a trapezoidal shape, which begins and ends at $J=0$ and ramps up and down over a finite time duration $\tau$:
\begin{equation}
f(t,T)=\begin{cases}
  t/\tau&0\le t\le\tau\\
  1&\tau<t\le T-\tau\\
  (T-t)/\tau&T-\tau<t \le T
\end{cases}\quad,\label{eq:finiterise2}
\end{equation}
and has total duration $T$.

We find the pulse sequence corresponding to a given $\tau>0$ value following the strategy presented in Ref.~\onlinecite{Wang.12}. Briefly, we replace each piece of the rotation in the no-rise-time case by one with the desired finite rise time, achieving the same target rotation at zeroth order and possessing the same number of free parameters to be determined by setting the total first-order error to zero. In Fig.~\ref{fig:finiterizeshape}, we present the representative pulse shape for $R(\hat{x}+\hat{z},\pi/2)$. Fig.~\ref{fig:finiterizeshape}(a) is the ideal square pulse with no rise time, while Fig.~\ref{fig:finiterizeshape}(b) shows the sequence for finite rise time $\tau=0.1$. In Fig.~\ref{fig:finiterizepara}, we present how the parameters of Fig.~\ref{fig:xpluszpulse} change if one still wants to correct error while having a finite rise time. As can be clearly seen from the figure, the parameters change very slightly. This means again that our method is robust, in the sense that for a non-rectangular pulse shape, one may find the desired solution through a simple local search around the solution we have for the ideal case. In practice, this can be done directly in experiments if an appropriate measure of fidelity is directly optimized.

\section{Two-qubit and multi-qubit operations}\label{sec:twoq}

\subsection{Two-qubit gates corrected through BB1 sequence}\label{sec:twoqBB1}

\begin{figure*}[]
    \centering
    \includegraphics[width=17cm, angle=0]{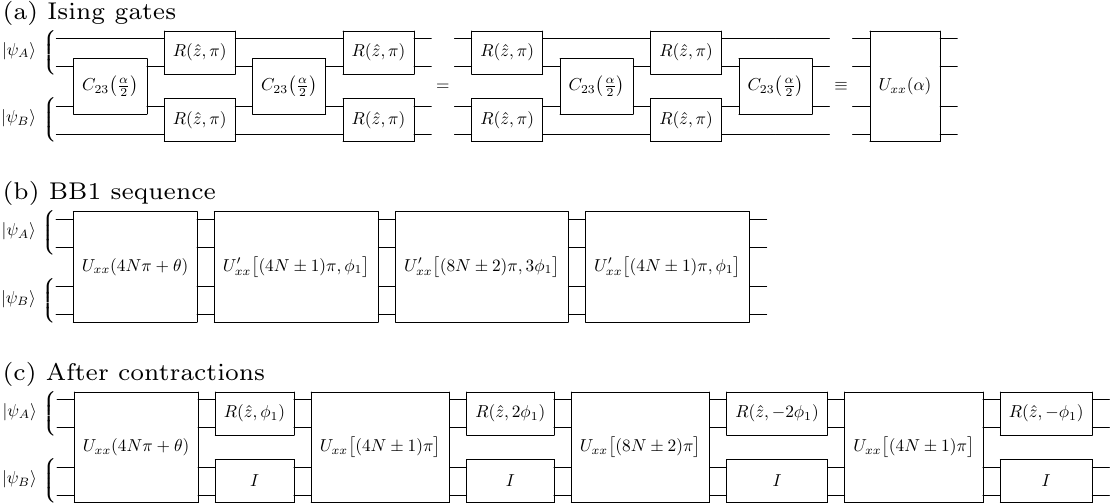}
    \caption{Quantum circuits for corrected two-qubit gates with the BB1 sequence. Each line represents a quantum dot,  linked to its neighboring dots by the exchange interaction.  A singlet-triplet qubit then is denoted as $|\psi_A\rangle$ and $|\psi_B\rangle$ by a pair of lines with corresponding electronic states within $S_z=0$ subspace. Panel (a) shows the Ising gate [Eq.~\eqref{eq:Klinovaja}]. Panel (b) shows the BB1 sequence [Eq.~\eqref{eq:BB1improvedtwoq}]. Here, $\phi_1$ is defined in Eq.~\eqref{eq:BB1improvedphi} if we assume $g(J)=J/\epsilon_0$. Panel (c) shows the result after contraction of $z$-rotations, and is identical to the sequence shown in Ref.~\onlinecite{Kestner.13} when taking $N=2$ and the plus sign in ``$\pm$''.}
    \label{fig:BB1}
\end{figure*}

In this section, we discuss two-qubit gates as well as how to perform a noise-resistant operation on a multi-qubit array. For this purpose we consider two neighboring qubits residing in four dots, with qubit $A$ residing in dots 1 and 2 and qubit $B$ residing in dots 3 and 4. To perform a two-qubit gate, one needs to couple electrons in dots 2 and 3, which in turn couples the two qubits. There are currently two ways to achieve this. One way is to use the capacitive coupling, as demonstrated experimentally in Ref.~\onlinecite{Shulman.12}. The other possibility is to use the exchange link between the neighboring two dots of the two double-dot system.\cite{Li.12,Klinovaja.12} The exchange coupling has the advantage that it is much stronger than the capacitive coupling, allowing for faster gates. We focus on the latter case in this work.

We first briefly explain how the exchange link between dots 2 and 3 can be combined with other single-qubit gates to achieve an entangled two-qubit gate, equivalent to {\sc cnot} up to single-qubit operations. Then we proceed to show how one can make this gate robust against noise, and how the entire sequence may be optimized.

The fundamental ingredient in performing a two-qubit gate with an exchange coupling is the Ising gate, labeled by $U_{xx}\left(\alpha\right)$, which is designed in such a way that in the absence of noise, leakage to the undesired states ${|\!\uparrow\uparrow\downarrow\downarrow\rangle}$ and ${|\!\downarrow\downarrow\uparrow\uparrow\rangle}$ is prevented, and one is left with a state-dependent phase due to the exchange pulse.\cite{Klinovaja.12} The pulse sequence reads
\begin{align}
U_{xx}\left(\alpha\right)&\equiv R^{\left(A\right)}\left(\hat{z},\pi\right) R^{\left(B\right)}\left(\hat{z},\pi\right) C_{23}\left(\frac{\alpha}{2}\right)\notag
\\
&\quad\times R^{\left(A\right)}\left(\hat{z},\pi\right) R^{\left(B\right)}\left(\hat{z},\pi\right) C_{23}\left(\frac{\alpha}{2}\right)\notag
\\
&= C_{23}\left(\frac{\alpha}{2}\right)R^{\left(A\right)}\left(\hat{z},\pi\right) R^{\left(B\right)}\left(\hat{z},\pi\right) \notag
\\
&\quad\times C_{23}\left(\frac{\alpha}{2}\right)R^{\left(A\right)}\left(\hat{z},\pi\right) R^{\left(B\right)}\left(\hat{z},\pi\right)\notag
\\
&= \exp\left(i\frac{\alpha}{2} \sigma_x\otimes\sigma_x\right) + \mathcal{O}\left(\delta h, \delta\epsilon\right),
\label{eq:Klinovaja}
\end{align}
and is also shown schematically in Fig.~\ref{fig:BB1}(a). Here, $R^{\left(A,B\right)}\left(\hat{z},\pi\right)$ denotes the {\sc swap} operation (i.e. ${|\!\uparrow\downarrow\rangle}\rightarrow{|\!\downarrow\uparrow\rangle}$) of qubits $A$ and $B$. $C_{23}\left(\alpha/2\right)$ denotes the application of a pulse to the inter-qubit exchange link $J_{23}$, such that for spins in dots 2 and 3, it acts as a $2\pi$ rotation (i.e., an identity) in the $S_z=0$ subspace, so that one may avoid swapping the spins in dots 2 and 3 (which would change ${|\!\uparrow\downarrow\uparrow\downarrow\rangle}$ to ${|\!\uparrow\uparrow\downarrow\downarrow\rangle}$), thereby preventing leakage. Its argument, $\alpha/2$, is fixed by $\int dt J_{23}\left(t\right) = \alpha/2$, corresponding to the desired relative phase to be acquired by two-qubit states. Since there are infinitely many ways to do an identity operation in the subspace of dots 2 and 3, one may choose one that has the correct pulse area to obtain the desired $\alpha$ by, for example, changing the axis of the $2\pi$ rotation. During a single application of $C_{23}\left(\alpha/2\right)$, the Overhauser fields also contribute to the phase accumulated by the two-qubit states, however, these may be removed by flipping all the spins in qubits $A$ and $B$  (the effect of the {\sc swap} gates) then applying the $C_{23}\left(\alpha/2\right)$ again, by which an equal amount of phase with opposite sign is accumulated through the Overhauser field. The {\sc swap} gates on qubits $A$ and $B$ are applied for this purpose. After the extra phase factor is canceled, the {\sc swap} gates must be applied once again to return the states to their original form carrying the desired phase factor. On the other hand, this entire sequence can also be reversed without making any difference in the absence of noise [cf. Eq.~\eqref{eq:Klinovaja} and Fig.~\ref{fig:BB1}(a)]. With an Ising gate handy, it is then very straightforward to convert it to any two-qubit gate.  Explicit formulae have been given in Refs.~\onlinecite{Klinovaja.12} and \onlinecite{Li.12} on how one may generate a {\sc cnot} gate from $U_{xx}(\alpha=\pi/2)$ and single-qubit operations.

Noise has two effects on the Ising gate. First, imperfect single-qubit rotations produce errors in the {\sc swap} gates and the $C_{23}$ gates, which, in the former cases contribute to decoherence and in the latter cases cause leakage out of the computational subspace in addition to decoherence. These errors can be corrected up to the first order by replacing all single-qubit gates by the {\sc supcode} sequences discussed in Sec.~\ref{sec:oneqrot}. What makes the noise correction difficult is that the charge noise makes $\int dt J_{23}\left(t\right)$, and subsequently the phase $\alpha$, erroneous. Worse, {\sc supcode} typically sweeps more than $14\pi$ around the Bloch sphere, during which a considerable amount of error may be accumulated. Since this problem is not solved by replacing all gates by {\sc supcode}, we must deal with it differently.

In Ref.~\onlinecite{Kestner.13} we propose a way to correct the phase error due to the charge noise by making an analogy to the BB1 sequence, which is known to correct the over-rotation error in NMR literature.\cite{Wimperis.94,Bando.13} In fact, although BB1 was originally proposed to correct over-rotation error for single-qubit gates, it was later realized \cite{Jones.03} that for the Ising two-qubit gate, a variant of the same sequence would work. We will therefore start with the single-qubit version of the BB1 sequence, and then verify whether that sequence can correct the over-rotation error as desired.

We start by defining a few notations. First, an $x$-rotation with an over-rotation error $\varepsilon$ is defined as
\begin{equation}
X(\varepsilon,\theta)=\exp\left[-i\sigma_x\frac{(1+\varepsilon)\theta}{2}\right]\label{eq:BB1Xgate}
\end{equation}
[We use a different notation than Eq.~\eqref{eq:UJphi} since here we have a different source of error.] Secondly, we tilt the entire frame around the $z$-axis by angle $\phi$:
\begin{equation}
X'(\varepsilon,\theta,\phi)=R(\hat{z},-\phi)X(\varepsilon,\theta)R(\hat{z},\phi)\label{eq:BB1Xprimegate}
\end{equation}
[note that $R(\hat{z},\pm\phi)$ here denotes ideal rotations].
The BB1 sequence is then based on the following identity:
\begin{align}
X'(\varepsilon,\pi,\phi_1)X'(\varepsilon,2\pi,3\phi_1)X'(\varepsilon,\pi,\phi_1)X(\varepsilon,\theta)\notag\\
=R(\hat{x},\theta) \left[I-\frac{i}{2}(\theta+4\pi\cos\phi_1)\varepsilon\right]+{\cal O}(\varepsilon^2)\label{eq:BB1basics}
\end{align}
and when
\begin{equation}
\phi_1=\pm\arccos\left(-\frac{\theta}{4\pi}\right)\label{eq:BB1basicsphi}
\end{equation}
Eq.~\eqref{eq:BB1basics} is robust against error at least up to the first order.

We may then make an analogy between $X(\varepsilon,\theta)$ and $U_{xx}(\alpha)$. Note that when making this direct analogy, we are already assuming that the overrotation error, $\int \delta J(t)$, is proportional to the rotation angle $\int J(t)$. This in turn means that we are assuming $g(J)=J/\epsilon_0$ here. In fact, our method works equally well when $g(J)$ assumes another form. We will postpone the discussion of this until after we have presented the SK1 version of the sequence in the next subsection, Sec.~\ref{sec:twoqSK1}. Here, for the convenience of presentation, we assume $g(J)=J/\epsilon_0$.

We must solve another problem before we can proceed. Eq.~\eqref{eq:BB1basicsphi} requires $-4\pi\le\theta\le4\pi$. Nevertheless, in our $U_{xx}(\alpha)$ gate, since we implement $C_{23}$ with {\sc supcode}, which sweeps more than $14\pi$ around the Bloch sphere, the integrated pulse area is typically larger than $4\pi$. Therefore we must generalize Eq.~\eqref{eq:BB1basics} to accommodate larger angles. To do this we observe that
\begin{align}
&X'\left[\varepsilon,(4N\pm1)\pi,\phi_1\right]X'\left[\varepsilon,(8N\pm2)\pi,3\phi_1\right]\notag\\
&\quad\cdot X'\left[\varepsilon,(4N\pm1)\pi,\phi_1\right]X(\varepsilon,4N\pi+\theta)\notag\\
&=R(\hat{x},\theta) \left\{I-\frac{i}{2}\left\{\theta+4\pi\left[N+(4N\pm1)\cos\phi_1\right]\right\}\varepsilon\right\}\notag\\
&\quad+{\cal O}(\varepsilon^2),\label{eq:BB1improved}
\end{align}
where $N$ is an integer. It is obvious that when we take $N=0$ and the plus sign in ``$\pm$'', Eq.~\eqref{eq:BB1improved} reduces to Eq.~\eqref{eq:BB1basics}. Now to make the first-order error vanish, we need
\begin{equation}
\phi_1=\pm\arccos\left[-\frac{4N\pi+\theta}{(16N\pm4)\pi}\right].\label{eq:BB1improvedphi}
\end{equation}
Here, by choosing an appropriate integer $N$, one may find real $\phi_1$ values for virtually any $\theta$.

To implement the sequence in the two-qubit scenario, we define a tilted version of the Ising gate $U_{xx}(\alpha)$, denoted by $U_{xx}'(\alpha,\phi)$: (note that both are subject to noise even though we did not explicitly indicate it)
\begin{align}\label{eq:tiltedKlinovaja}
U_{xx}'\left(\alpha,\phi\right) &\equiv  R^{\left(A\right)}\left(\hat{z},-\phi\right) U_{xx}\left(\alpha\right) R^{\left(A\right)}\left(\hat{z},\phi\right)\notag\\
&=[R\left(\hat{z},-\phi\right)\otimes I]U_{xx}\left(\alpha\right)[R\left(\hat{z},\phi\right)\otimes I]
\end{align}
Here, the $z$-rotations are only done on qubit $A$. While there is no doubt that such a $z$-rotation on qubit $A$ needs to be done with {\sc supcode}, qubit $B$ also has to undergo a {\sc supcode} identity operation corrected against noise up to the first order, with the same time duration as the operation on qubit $A$. We will discuss this in more detail in Sec.~\ref{sec:ident}.

By making direct analogy to Eq.~\eqref{eq:BB1improved}, one may easily verify that when $\phi_1$ is chosen as prescribed in Eq.~\eqref{eq:BB1improvedphi}, the sequence
\begin{align}
&U_{xx}'\left[(4N\pm1)\pi,\phi_1\right]U_{xx}'\left[(8N\pm2)\pi,3\phi_1\right]\notag\\
&\quad\cdot U_{xx}'\left[(4N\pm1)\pi,\phi_1\right]U_{xx}(4N\pi+\theta)\notag\\
&=\exp\left(i\frac{\theta}{2} \sigma_x\otimes\sigma_x\right) + {\cal O}\left[\left(\delta h + \delta\epsilon \right)^2\right],\label{eq:BB1improvedtwoq}
\end{align}
is immune to noise up to the leading order. This sequence is shown in Fig.~\ref{fig:BB1}(b). [Note that in the figure, time flows from left to right, while on the left hand side of Eq.~\eqref{eq:BB1improvedtwoq} from right to left.]

There are some trivial optimizations. For two consecutive $U_{xx}'$ operations, the $z$-rotations in the middle of them can be contracted, so one is to perform one instance of {\sc supcode} for them rather than two. For example, $U_{xx}'\left[(4N\pm1)\pi,\phi_1\right]$ has $R(\hat{z},\phi)\otimes I$ on its right, while $U_{xx}'\left[(8N\pm2)\pi,3\phi_1\right]$ has $R(\hat{z},-3\phi)\otimes I$ on its left. Then, when they are applied back-to-back, the $z$-rotations can be contracted as $R(\hat{z},-2\phi)\otimes I$. The sequence after such contraction is shown in Fig.~\ref{fig:BB1}(c). If we take $N=2$, $\theta=\pi/2$, and choose the plus sign in ``$\pm$'', we have exactly the same sequence as presented in Ref.~\onlinecite{Kestner.13}.

Obviously, the cost of being immune to noise is a much longer gate time. For the sequence shown in  Fig.~\ref{fig:BB1}(c), 12 $z$-rotations and eight $C_{23}$ gates are required, each of which requires roughly $18\pi$ of rotation around the Bloch sphere. Therefore the total length of the gate, in terms of the angle swept, is around $360\pi$. We therefore are interested in further optimizing the corrected Ising gate.

\subsection{Two-qubit gates corrected through SK1 sequence}\label{sec:twoqSK1}

\begin{figure*}[]
    \centering
    \includegraphics[width=17cm, angle=0]{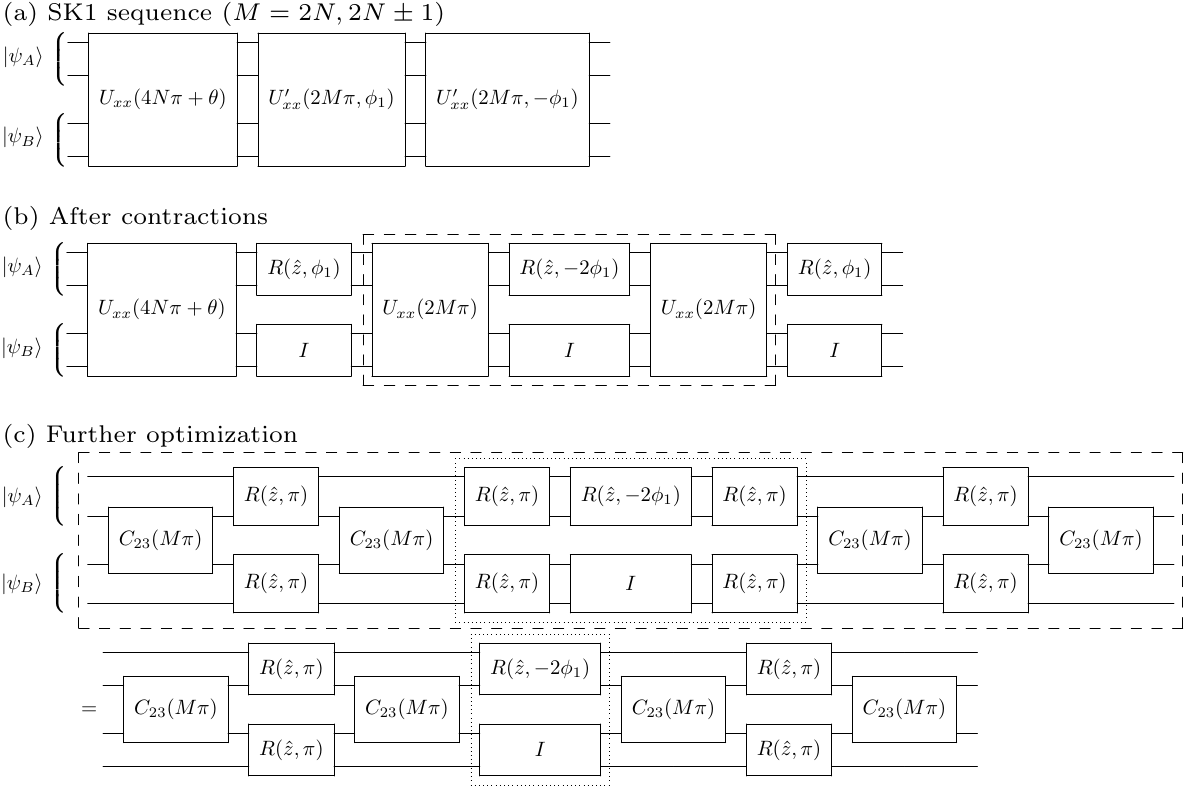}
    \caption{Quantum circuits for corrected two-qubit gates with the SK1 sequence. Panel (a) shows the SK1 sequence [Eq.~\eqref{eq:SK1improvedtwoq}], with $\phi_1$ defined in Eq.~\eqref{eq:SK1improvedphi} if we assume $g(J)=J/\epsilon_0$. Panel (b) shows the sequence after contractions are performed as in Fig.~\ref{fig:BB1}(c). Several gates are enclosed by the dashed frame, which are further optimized as explained in the main text [Eq.~\eqref{eq:SK1furtheroptimization}], and shown in panel(c).}
    \label{fig:SK1}
\end{figure*}

A $25\%$ reduction in the length of the gate can be made by using the SK1 sequence instead of BB1. The reason is simply that SK1 requires three components [$X$ and $X'$ defined in Eqs.~\eqref{eq:BB1Xgate} and \eqref{eq:BB1Xprimegate}], while BB1 requires four.

With functions $X$ and $X'$ already defined in Eqs.~\eqref{eq:BB1Xgate} and \eqref{eq:BB1Xprimegate}, we write the identity based on which the SK1 sequence is defined, as
\begin{align}
&\quad X'(\varepsilon,2\pi,-\phi_1)X'(\varepsilon,2\pi,\phi_1)X(\varepsilon,\theta)\notag\\
&=R(\hat{x},\theta) \left[I-\frac{i}{2}(\theta+4\pi\cos\phi_1)\varepsilon\right]+{\cal O}(\varepsilon^2)\label{eq:SK1basics}
\end{align}
and when $\phi_1$ is given by Eq.~\eqref{eq:BB1basicsphi}, the first order error vanishes.

Similarly to the BB1 case, we need to generalize Eq.~\eqref{eq:SK1basics} so that larger values of $\theta$ are allowed. Straightforward algebra shows
\begin{align}
&X'\left(\varepsilon,2M\pi,-\phi_1\right)X'\left(\varepsilon,2M\pi,\phi_1\right)X(\varepsilon,4N\pi+\theta)\notag\\
&=R(\hat{x},\theta) \left\{I-\frac{i}{2}\left\{\theta+4\pi\left[N+M\cos\phi_1\right]\right\}\varepsilon\right\}\notag\\
&\quad+{\cal O}(\varepsilon^2),\label{eq:SK1improved}
\end{align}
where $M$ can take the values $\{2N\pm1,2N\}$ where $N$ is an integer. To make the first-order error vanish, we require
\begin{equation}
\phi_1=\pm\arccos\left(-\frac{4N\pi+\theta}{4M\pi}\right).\label{eq:SK1improvedphi}
\end{equation}

 Then, with $U_{xx}'(\alpha,\phi)$ defined in Eq.~\eqref{eq:tiltedKlinovaja}, we write the SK1 sequence for the two-qubit gate as
\begin{align}
&U_{xx}'\left(2M\pi,-\phi_1\right)U_{xx}'\left(2M\pi,\phi_1\right)U_{xx}(4N\pi+\theta)\notag\\
&=\exp\left(i\frac{\theta}{2} \sigma_x\otimes\sigma_x\right) + {\cal O}\left[\left(\delta h + \delta\epsilon \right)^2\right].\label{eq:SK1improvedtwoq}
\end{align}
It is straightforward to verify that the sequence is indeed immune to noise up to the leading order if $\phi_1$ is chosen according to Eq.~\eqref{eq:SK1improvedphi}. This sequence is shown in Fig.~\ref{fig:SK1}(a). We can then do contractions of $z$-rotation gates as we did in Fig.~\ref{fig:BB1}(c). The resulting sequence is shown in Fig.~\ref{fig:SK1}(b).

Further optimization may be made by noting that in the definition of $U_{xx}$, the order of the sequence may be reversed [Fig.~\ref{fig:BB1}(a)]. We choose the partial sequence $U_{xx}(2M\pi)\cdot [R(\hat{z},-2\phi_1)\otimes I]\cdot U_{xx}(2M\pi)$, as enclosed by the dashed frame in Fig.~\ref{fig:SK1}(b), and expand the  two $U_{xx}(2M\pi)$ gates in such a way that their end parts, the $R(\hat{z},\pi)\otimes R(\hat{z},\pi)$ gates, are leaning toward each other. Then one may simply do the contraction
\begin{align}
&[R(\hat{z},\pi)\otimes R(\hat{z},\pi)]\cdot[R(\hat{z},-2\phi_1)\otimes I]\cdot[R(\hat{z},\pi)\otimes R(\hat{z},\pi)]\notag\\
&\quad=R(\hat{z},-2\phi_1)\otimes I\label{eq:SK1furtheroptimization}
\end{align}
as shown in the dotted frame of Fig.~\ref{fig:SK1}(c).

For the sequence shown in  Fig.~\ref{fig:SK1}, seven $z$-rotations and six $C_{23}$ gates are required. If we assume that each gate requires roughly $18\pi$ of rotation around the Bloch sphere, the total length of the gate, in terms of the angle swept, is around $230\pi$, a 35\% reduction from the BB1 sequence.

The above discussions of the BB1 and SK1 sequences have assumed $g(J)=J/\epsilon_0$, which means that the over-rotation error is proportional to the angle $\theta$ rotated [cf. Eq.~\eqref{eq:BB1Xgate}]. In the general case, we may revise Eq.~\eqref{eq:BB1Xgate} with the over-rotation error acquiring a factor dependent on $\theta$:
\begin{equation}
\widetilde{X}(\varepsilon,\theta)=\exp\left\{-i\sigma_x\frac{\{1+[1-\lambda(\theta)]\varepsilon\}\theta}{2}\right\}.\label{eq:BB1revisedXgate}
\end{equation}

We shall demonstrate how our method works for the SK1 sequence, but application to BB1 sequence is conceptually the same.
Corresponding to Eq.~\eqref{eq:BB1revisedXgate}, in Eq.~\eqref{eq:SK1improved} we need the function $\widetilde{X}$ for two rotation angles, $2M\pi$ and $4N\pi+\theta$. This means that $\lambda(\theta)$ in Eq.~\eqref{eq:BB1revisedXgate} may take two possibly different values. We denote
$\lambda(4N\pi+\theta)\equiv\lambda_1$ and $\lambda(2M\pi)\equiv\lambda_2$.
Similarly to Eq.~\eqref{eq:BB1Xprimegate}, we define
\begin{equation}
\widetilde{X}'(\varepsilon,\theta,\phi)=R(\hat{z},-\phi)\widetilde{X}(\varepsilon,\theta)R(\hat{z},\phi)\label{eq:BB1Xprimegaterevised}
\end{equation}
and Eq.~\eqref{eq:SK1improved} must be correspondingly revised as
\begin{align}
&\widetilde{X}'\left(\varepsilon,2M\pi,-\phi_1\right)\widetilde{X}'\left(\varepsilon,2M\pi,\phi_1\right)\widetilde{X}(\varepsilon,4N\pi+\theta)\notag\\
&=R(\hat{x},\theta)\notag\\
&\times\left\{I-\frac{i}{2}\left[(1-\lambda_1)(\theta+4N\pi)+(1-\lambda_2)4M\pi\cos\phi_1\right]\varepsilon\right\}\notag\\
&\quad+{\cal O}(\varepsilon^2),\label{eq:SK1improvedrevised}
\end{align}
and $\phi_1$ has to satisfy
\begin{equation}
\phi_1=\pm\arccos\left[-\frac{(4N\pi+\theta)(1-\lambda_1)}{4M\pi(1-\lambda_2)}\right].\label{eq:SK1improvedphirevised}
\end{equation}
to make the first-order error vanish.

Therefore under a general scenario with an arbitrary dependence of $J$ on detuning, our sequence will work perfectly as long as one chooses the $\phi_1$ value as in Eq.~\eqref{eq:SK1improvedphirevised}. Here we made no assumption about the precise form of $g(J)$: the only important thing is that $g(J)$ has to be known and the values of $\lambda_1$ and $\lambda_2$ can be calculated.

\subsection{Manipulation of a multi-qubit system and the buffering identity operation}\label{sec:ident}

Since we now have corrected single-qubit and  two-qubit gates, arbitrary multi-qubit circuits immune to noise can be performed in a similar manner as shown in Fig.~\ref{fig:BB1} and Fig.~\ref{fig:SK1}. There remains one more component essential for implementing a multi-qubit circuit: a variable-time identity operation. In fact, the identity operation plays important roles in several parts of our pulse sequence, which we explain in detail below.

First, as can be seen from Fig.~\ref{fig:BB1} and Fig.~\ref{fig:SK1}, when qubit $A$ is undergoing a single-qubit operation, for example a $z$-rotation, qubit $B$ has to undergo a corrected identity operation. One cannot simply do nothing on qubit $B$ because the constant presence of the Overhauser field required to access the $x$-axis rotation would lead the qubit states to stray undesirably, and the situation is made worse by the presence of noise. Therefore it is necessary for qubit $B$ to undergo a corrected identity operation which has the same time duration as the operation performed on qubit $A$, namely they must both end at the same time and proceed to the next operation. The same holds for multi-qubit gates: when several qubits as part of a qubit array are performing certain operations, all remaining qubits must perform identity operations, and these operations should all have the same time duration. If the operations on two qubits have different time durations (say $t_1$ and $t_2$), then one must supplement those operations with identity operations with time durations $T-t_1$ and $T-t_2$ to make them end at the same time, while the remaining qubits must also end their respective operations at time $T$. This is necessary to keep the entire system immune to noise to the leading order. For example, in the Ising gate as shown in Fig.~\ref{fig:BB1}(a), both qubits $A$ and $B$ have to perform a {\sc swap} operation, $R(\hat{z},\pi)$. They would automatically span the same time if the Overhauser fields for qubits $A$ and $B$ are identical. However when the Overhauser fields are different, the two operations would end at different times and one must ``buffer'' them by identities as explained above.

Secondly, the identity operation is also fundamental for two-qubit gates: it is an essential ingredient for performing $C_{23}\left(\frac{\alpha}{2}\right)$ [cf. Eq.~\eqref{eq:Klinovaja} and Fig.~\ref{fig:BB1}(a)]. In the definition of $C_{23}$,  the argument $\alpha/2$ is equal to $\int dt J_{23}\left(t\right)$, which then directly translates to the resulting Ising gate $U_{xx}(\alpha)$. Here, $J_{23}(t)$ is a composite pulse implementing an identity operation in the $S_z=0$ subspace of dots 2 and 3, therefore generating $U_{xx}$ for a certain value of $\alpha$ amounts to doing an identity operation with $\int dt J\left(t\right)$ matching a predetermined value.

The above discussion implies that we need a family of corrected identity operations which can generate a broad range of time durations as well as values of $\int dt J\left(t\right)$. To accomplish this, we employ an additional degree of freedom in the discussion of Sec.~\ref{sec:onepiece}, the exchange interaction $J$. Note that for each value of $J$, one can always perform a corrected $2\pi$ rotation around $\hat{x}+J\hat{z}$ with a certain time $T$ and value of
 $\int dt J\left(t\right)$. When $J$ is changed between $0$ and $J_{\rm max}$, $T$ and  $\int dt J\left(t\right)$ would also change, covering certain ranges.

We have found such an identity as a level-6 one, defined as
\begin{align}
\begin{split}
&U(J,2\pi)
U(j_5,\pi)
U(j_4,\pi)
U(j_3,\pi)
U(j_2,\pi)\\
&\times
U(j_1,\pi)
U(j_0,4\pi)
U(j_1,\pi)
U(j_2,\pi)
U(j_3,\pi)\\
&\times
U(j_4,\pi)
U(j_5,\pi)
U(J,2\pi)
\end{split}\label{eq:varidentity}
\end{align}
Since the sequence is symmetric, we only need four unknowns. We then choose $j_2=j_4=0$, and use $J$ as the ``tunable knob'': for each given value of $J$, we solve for physical solutions of $j_0$, $j_1$, $j_3$, $j_5$, and record the time duration and $\int dt J(t)$. We have found that the pulse sequence of Eq.~\eqref{eq:varidentity} generates identity operations with time duration $T_f$ between $22$ and $48$. By duplicating this identity, one can obtain corrected identities spanning any time for $T_f>22$. These identity operations can also be used in the construction of the two-qubit gates discussed in previous sections. This pulse sequence generates values of $\int dt J\left(t\right)$ between $10$ and $20$ (corresponding to $20<\alpha<40$). In Ref.~\onlinecite{Kestner.13}, we have used this sequence to generate  $U_{xx}(17\pi/2)$ and $U_{xx}(9\pi)$.

\section{Randomized Benchmarking}\label{sec:benchmarking}
\begin{figure*}[p]
  \includegraphics{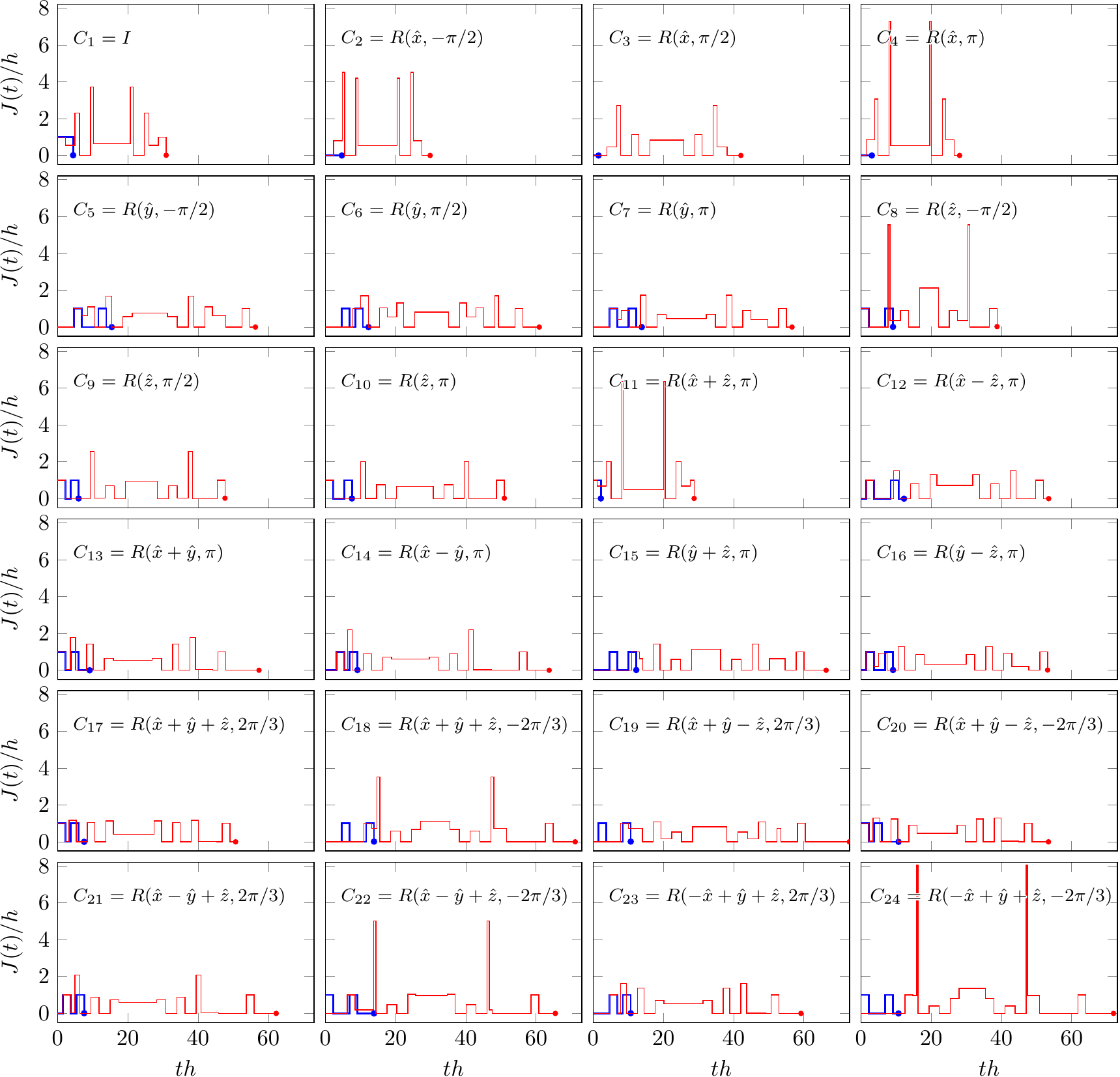}
  \caption{\label{fig:fignaiv}(Color online) Implementations of all $24$ single-qubit Clifford gates. Thick blue: na\"ive implementation; Fine red: \textsc{supcode} implementation.}
\end{figure*}
\begin{figure}[tbp]\raggedleft
  \includegraphics{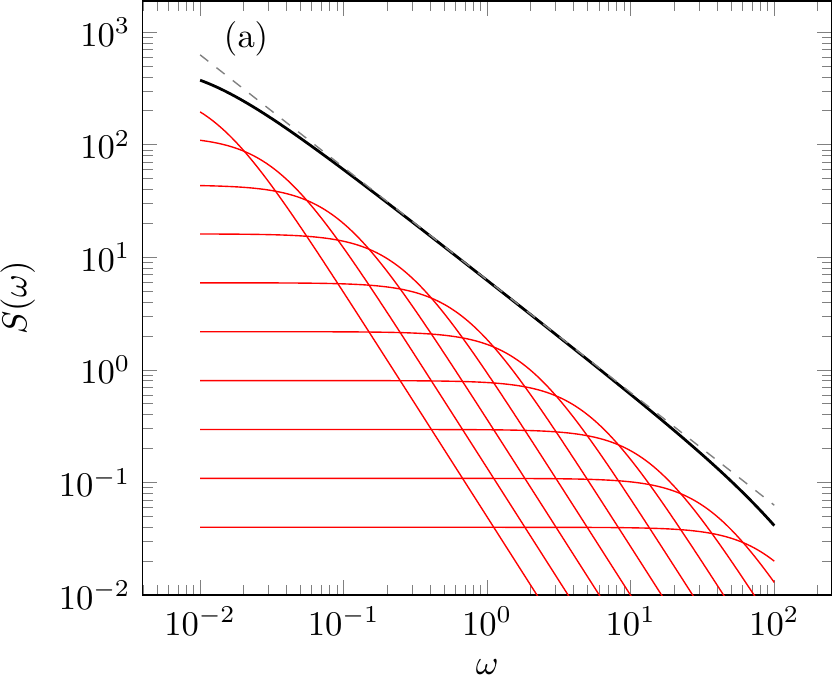}
  \includegraphics{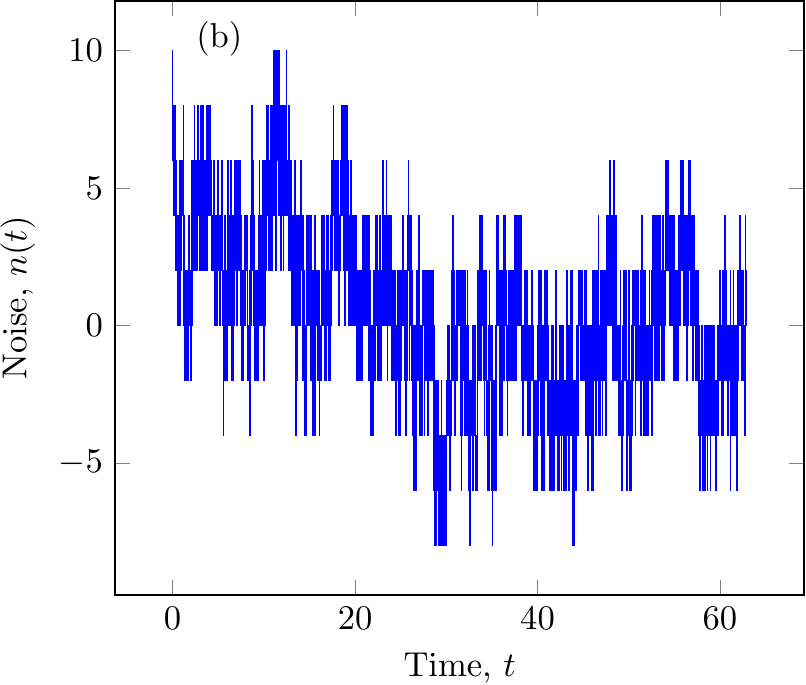}
  \caption{\label{fig:figoverf}(Color online) Finite-bandwidth approximation of $1/\omega$ noise via sum of RTS\@. (a)~Thin red: power spectra of individual RTSs; thick black: sum of RTS; dashed: ideal $1/\omega$. (b)~A specific sample of such noise drawn from this distribution.} \end{figure}
\begin{figure}[tbp]
  \includegraphics{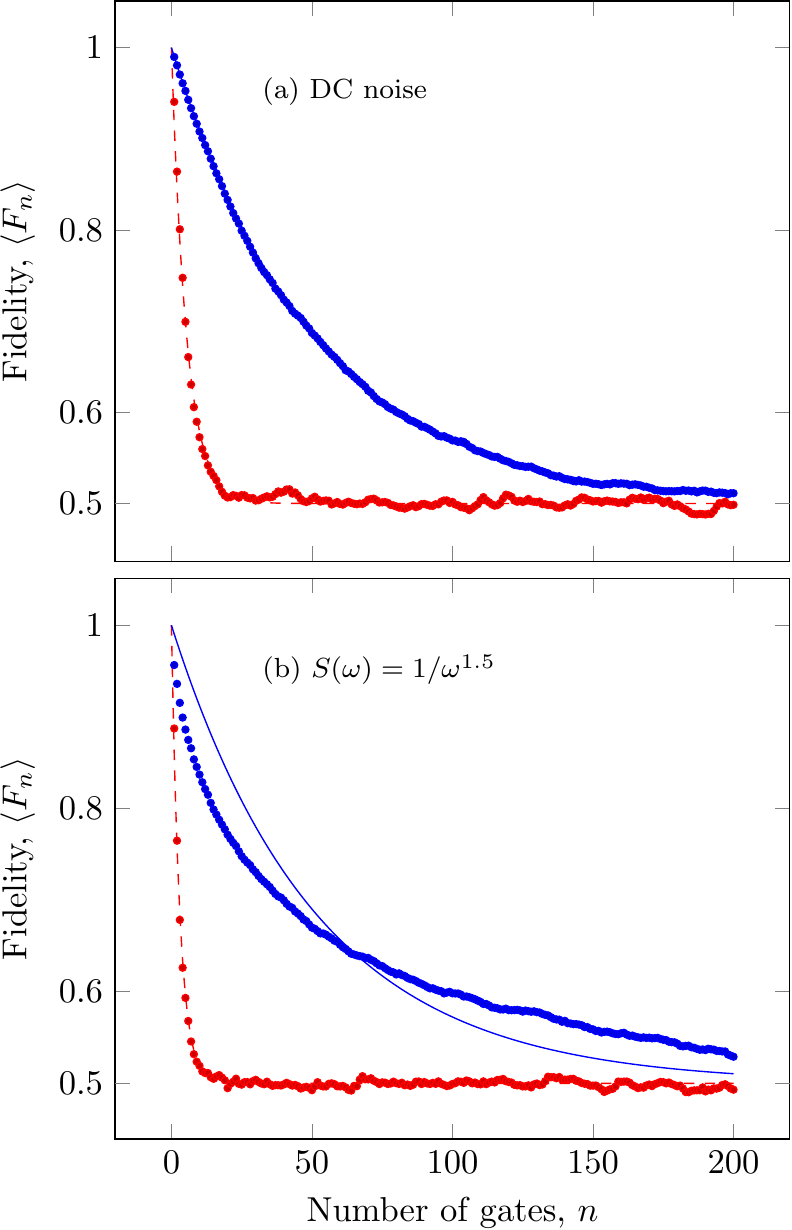}
  \caption{\label{fig:figDCexp}(Color online) Fidelity vs number of gates. (a)~DC noise, (b)~$1/\omega^{1.5}$ noise. Red/dashed: na\"ive Clifford implementation; blue/solid: \textsc{supcode} Clifford implementation. Points are from RB simulation and curves are fits to $(1+e^{-\gamma n})/2$. Note that the exponential decay model does not fully describe the data in part (b).}
\end{figure}
\begin{figure}[tbp]
  \includegraphics{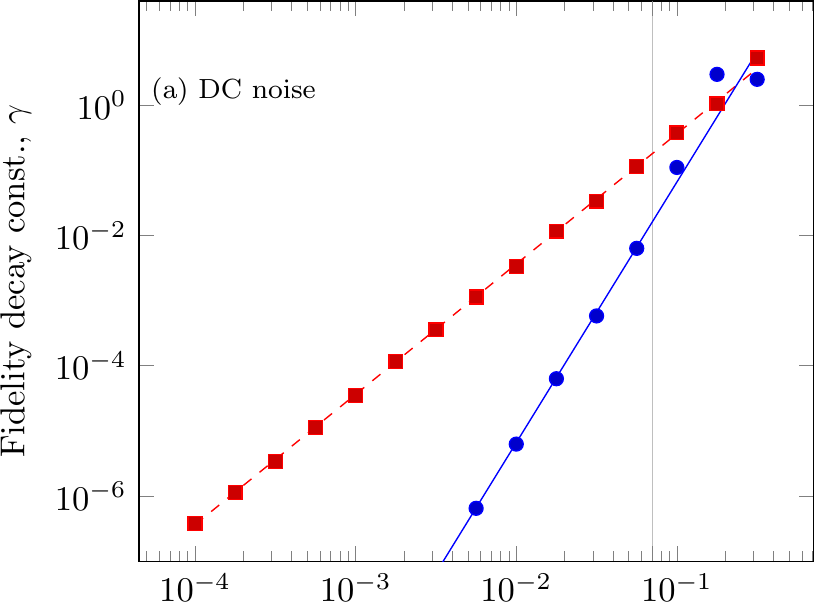}
  \includegraphics{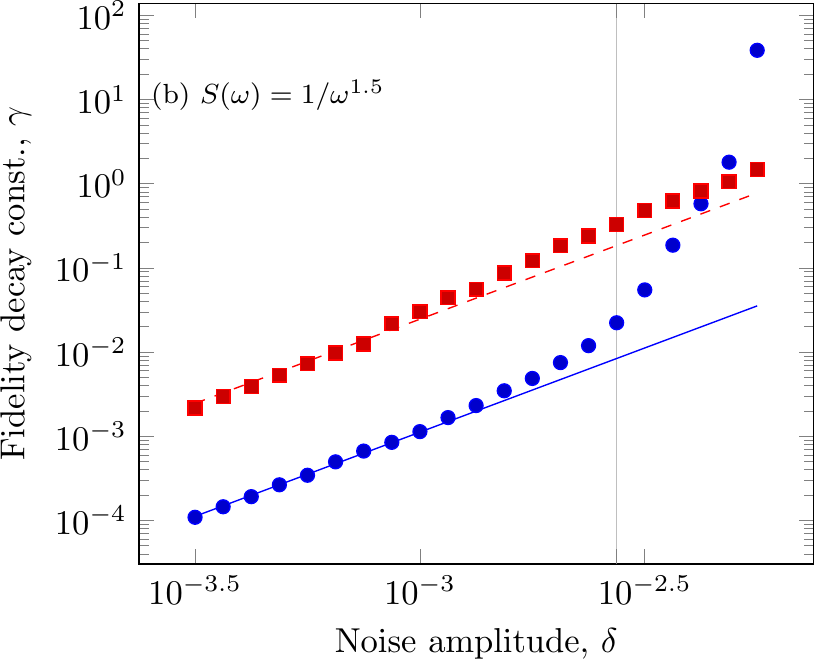}
  \caption{\label{fig:figdjdh}(Color online) Fidelity decay constant $\gamma$ vs noise amplitude. (a)~DC noise, (b)~$1/\omega^{1.5}$ noise. Red/dashed: naive Clifford implementation; blue/solid: \textsc{supcode} Clifford implementation. Points are from RB simulation, curves are proportional to  $\delta^2$ and $\delta^4$. For DC noise, as expected, the lowest-order contribution of the noise is canceled by \textsc{supcode}, leaving a residual effect $\mathcal{O}(\delta^4)$. For the AC noise, the improvement from \textsc{supcode} saturates at approximately 10-fold reduction in $\gamma$. Vertical lines indicate the values of $\delta$ used in Fig.~\ref{fig:figDCexp}.}
\end{figure}
\begin{figure}[tbp]
  \includegraphics{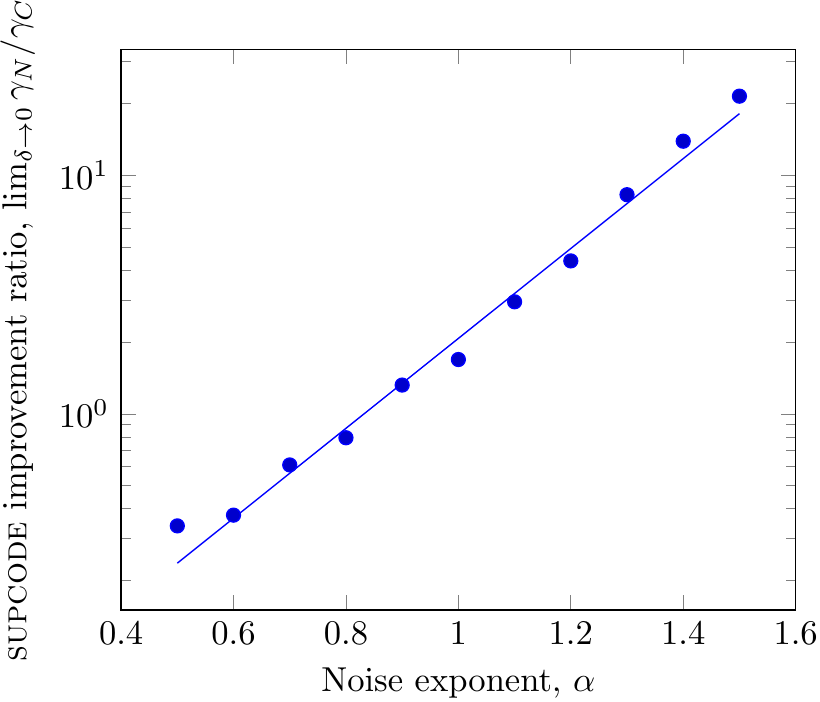}
  \caption{\label{fig:figratio}(Color online) Asymptotic improvement ratio of \textsc{supcode} vs na\"ive pulses, for $1/\omega^\alpha$ noise. The line is $2\times 76^\alpha$}
\end{figure}
The \textsc{supcode} sequences cancel lowest-order effects of static (DC) noise, but they
will not function in the opposite limit, of completely white
noise. In reality, we expect the noise spectrum to be of an intermediate
$\omega^{-\alpha}$ form, where recent experimental work puts $\alpha\simeq2.5$ for nuclear spin fluctuations \cite{Medford.12,rudner_nuclear_2011}), $\alpha=0.7$ for charge noise.\cite{Dial.13}

For such ``colored'' noise, the slow correlations mean that it is not
necessarily possible to predict the fidelity of a quantum algorithm
involving a sequence of gates from looking at the performance of the
individual gates within that sequence. A powerful technique for
investigating the fidelity of pulse sequences exists in the form of
Randomized Benchmarking~(RB) (a nice theoretical overview is given in Ref.~\onlinecite{magesan_characterizing_2012}). The crucial insight of RB is that instead of
investigating arbitrary gates, we may restrict ourselves to a finite
subset, the Clifford group. This means that we need only
produce a finite set of corrected gates. Also, we can efficiently
calculate the effect of any arbitrary sequence of ideal Clifford gates
acting on a state. Additionally, after any arbitrary sequence of
Clifford gates applied to the system ground state $|0\rangle$, only 1 additional,
efficiently-calculable, Clifford gate is required to rotate the resulting
state into the standard $|0\rangle$-$|1\rangle$ measurement basis. This last property
is crucial for the experimental implementation of RB, allowing errors
in the Clifford gates to be determined independently of any errors in
state preparation and measurement (SPAM).

We have investigated the theoretical performance of our single-qubit gates
using a numerical simulation of Randomized Benchmarking. We denote the
set of 24 single-qubit Cliffords as $\{C_1,...,C_{24}\}$, and similarly denote
a pulse implementation of the group (which may be a na\"ive
uncorrected implementation, or one of our corrected \textsc{supcode} composite
pulse implementations as given explicitly in Tables \ref{tab:numericonepiece}~to~\ref{tab:genpulsepara} and shown graphically in Fig.~\ref{fig:fignaiv}) as $\{\tilde{C}_i\}$.

We can calculate the expected fidelity of an
implementation of a length-$n$ sequence of Cliffords as $\langle F_n\rangle = \langle F(C_{j_1}
C_{j_2}\cdots C_{j_n}, \tilde{C}_{j_1}
\tilde{C}_{j_2}\cdots \tilde{C}_{j_n}) \rangle$ where $F(A,B)$ denotes the
fidelity between unitaries $A$ and $B$, and the bracket $\langle\cdot\rangle$ represents averaging
over both the choice of random Clifford elements $j_1,j_2,...,j_n$ distributed
uniformly and independently over $\{1,2,\ldots,24\}$ and also averaging over
realizations of the charge and magnetic field noise, parameterized by an amplitude $\delta$.

We generate ``$1/f$'' noise
realizations via a weighted sum of Random Telegraph Signals~(RTSs)\cite{kogan},
resulting in noise that approximates a desired $\omega^{-\alpha}$ power spectrum over
a wide range in $\omega$, as shown in Fig.~\ref{fig:figoverf}. We choose the low-frequency cutoff such that the slowest RTS has time constant $\tau_\text{max}=10^4/h$ and the high-frequency cutoff from $\tau_\text{min}=1/h$. One interpretation of a low-frequency cutoff is that it corresponds to the experimentalist making a calibration of $h$ and $J$ on a timescale of $\tau_\text{max}$ prior to a given benchmarking run. As such, our choice of $\tau_\text{max}$ minimizes the relative improvement due to \textsc{supcode}, since it corresponds to calibrating out $\delta J$ and $\delta h$ about as quickly as is reasonable to imagine: more usually $\tau_\text{max}$ will be on the order of minutes or hours ($\tau_\text{max}\simeq 10^{11}/h$) leading to a much larger DC component of the noise and correspondingly better performance of \textsc{supcode} compared to na\"ive pulses. Our high-frequency cutoff is on the order of the shortest pulses of our sequences, such that all higher frequencies are effectively ``white'': extending the cutoff towards higher frequencies should be equivalent to adding a white noise background that will affect the na\"ive and corrected pulses similarly, depending only on their total duration.

Both because the na\"ive and corrected pulse sequences are built from
piecewise-constant pulses and because the noise realizations are also
piecewise-constant, the system evolution can be efficiently calculated
as a product of matrix exponentials. This gives an efficient
calculation of the expected fidelity $\langle F_n\rangle$. We proceed in the standard
way for RB by fitting $\langle F_n\rangle$ for differing $n=0,...,N$ to a decaying
exponential function
$\langle F_n\rangle = (1 + e^{-\gamma n})/2$,
where unlike in the case of experimental RB we are able to avoid
fitting an overall scaling factor, due to absence of SPAM errors for
this numerical simulation. Due to the non-Markovian form of the noise,
$\langle F_n\rangle$ is not necessarily expected to have exactly exponential form,
and indeed we do observe a deviation from the exponential in Fig.~\ref{fig:figDCexp}.
Nevertheless we use the fitted $\gamma$ [which can be related to an
error-per-gate (EPG)\cite{magesan_characterizing_2012}] to summarize the performance of a particular
implementation of the Clifford group under a particular noise
distribution.

When the strength of the noise is reduced, we find that,
as expected, the EPG of a \textsc{supcode} Clifford implementation falls more
steeply than for a na\"ive implementation (see Fig.~\ref{fig:figdjdh}). For static noise, the $\gamma$ for \textsc{supcode} is $\mathcal{O}(\delta^4)$ order in the
noise strength, $\delta$, compared to $\mathcal{O}(\delta^2)$ order for the na\"ive implementation,
allowing the \textsc{supcode} to perform arbitrarily better than the na\"ive
sequence, if the noise can be reduced sufficiently. However, for
colored noise, the ratio of the na\"ive $\gamma_N$ to the \textsc{supcode} $\gamma_C$
saturates to a finite value, $r$, in the limit that the noise is reduced
toward zero, $r=\lim_{\delta\to0}\gamma_N/\gamma_C$. Thus, there is a maximum improvement that is possible for
 \textsc{supcode}. We find that this ratio is a strong function of the exponent $\alpha$ of the noise distribution,
and over the range $0.5<\alpha<1.5$ it fits well to an exponential
function $2\times p^{\alpha-1}$ (see Fig.~\ref{fig:figratio}, where $p=76$).  (To study $\alpha$ much outside
this range, we would need to use a different process to generate the
noise.) The specific value of the base varies, $20\lesssim p\lesssim 80$, when sweeping the low- and high-frequency cutoffs of the noise spectrum over a factor of 10, but the sensitivity to $\alpha$ remains. Based on this empirical result and the experimental estimates of $\alpha$, it seems that \textsc{supcode} should perform
extremely well against magnetic field noise, but have more limited
success against charge noise. This assumes the experimental estimates of
$\alpha$ for these noises turn out to hold true, and comes with the caveat that a sum of RTSs cannot reproduce a noise spectrum with $\alpha>2$ where a spin-diffusion model is more physically realistic. A future variant of
\textsc{supcode} might trade a fraction of the performance against field noise for
improved performance against charge noise.

Our numerical RB technique can be extended in a straightforward, if
tedious, fashion to investigate 2-qubit sequences. We have only
considered the case where the magnetic field noise and charge noise
are of similar magnitude, have the same $\alpha$, and are generated
independently: it will be interesting to relax some of these
constraints. In particular it could be interesting to examine the
effect of correlated noises, and it may be possible to construct
families of pulse sequences that sacrifice some performance on general
independent noise in favor of performance on correlated noise.
Another open question relates to the failure of the Gaussian
approximation for colored noise --- the noise is not only
characterized by the power spectrum, but also by the microscopic
structure of the environment. For example, rather than our weighted
sum of RTSs, modeling the case where the noise is due to a collection
of two-level fluctuators with random switching rates, the same noise
spectrum could arise from a single fluctuator with an undetermined
switching rate. Due to the failure of the Gaussian approximation,
these different environments may cause different behavior (see, for example Ref.~\onlinecite{benedetti_dynamics_2013}). Our numerical technique can be extended
to investigate the behavior of our gates under such different
environments.

\section{Conclusion}\label{sec:conclusion}

In conclusion, we have shown that our protocol\cite{Wang.12,Kestner.13} for performing robust quantum control of semiconductor spin qubits, {\sc supcode}, can be extended to incorporate the numerous complications inherent in a real quantum device without compromising any of its error-suppressing capabilities. We have shown that this is true for both the full range of single-qubit operations as well as for an entangling two-qubit gate, demonstrating that noise-resistant universal quantum control can be achieved in actual experiments. In the case of the two-qubit gate, we have also explained how the gate operation time can be substantially reduced compared to earlier work, constituting a crucial step toward experimental implementation. In addition, we have provided a randomized benchmarking for our proposed gate control operations. Below, we summarize our main findings regarding each of these points.

The most important message of this work is that the applicability of {\sc supcode} is not in any way diminished when various experimental complications are taken into account. One such complication stems from the dependence of the exchange coupling on the detuning. This dependence varies from sample to sample and has a large impact on the effect of charge noise on the qubit, so it is therefore important that schemes to combat charge noise such as {\sc supcode} are able to incorporate this dependence into their functionality. In our earlier work on {\sc supcode}, as well as in other theoretical and experimental works, a simple model in which the exchange coupling is assumed to increase exponentially with the detuning was used. While this assumption can greatly simplify the theoretical analysis, it also raises the question of whether the efficacy of {\sc supcode} depends on this assumption. Here, we have explicitly shown that this is not the case, and that {\sc supcode} remains equally effective for other models of the exchange-coupling dependence on detuning. In fact, for a general model, we have seen that one simply needs to adjust the form of the coupled nonlinear equations and then follow the standard procedure to solve them to obtain error-suppressing pulse sequences. We demonstrated this fact explicitly for two alternative choices of the exchange coupling function and showed that numerical solutions can still be found. Furthermore, we have shown that these results hold for both the single and two-qubit gates.

A second complication that arises in real experiments is that pulses cannot be made perfectly square; instead they necessarily contain a finite rise time during which the exchange coupling switches between zero and non-zero values. Replacing the perfect square with a trapezoidal model for the pulses, we showed that a finite rise time would merely translate to rather small shifts in the pulse parameters relative to the values obtained for square pulses. We further showed that it is generally the case that one can start with the pulse parameters found assuming perfectly square pulses, and then optimize around these values to obtain noise-resistant sequences of pulses with finite rise times. The fact that finite rise times do not lead to a substantial change in the parameters means that such a search remains local in parameter space and is relatively easy to perform.

A third experimental reality that our earlier works on {\sc supcode} did not account for is the fact that the noise is not truly static, exhibiting some variation on longer time scales. To address the importance of this effect on the operation of {\sc supcode}, we presented a complete randomized benchmarking analysis showing that this indeed puts some limitations on the performance of {\sc supcode}.

To make {\sc supcode} experimentally feasible, it is not only important to account for the issues that arise in real physical systems, but it is also crucial to shorten the total gate operation times as much as possible. In particular, we showed that the length of the corrected two-qubit gate presented in Ref.~\onlinecite{Kestner.13} can be significantly reduced by about 35\%. This large reduction is made possible by replacing the BB1 sequence with a generalized SK1 sequence, in conjunction with some additional optimizations to the sequence. It is, in principle, likely that the pulse sequence can be shortened further through extensive numerical searches for better optimization, but given that the pulse sequences proposed in this work are already short enough for laboratory implementations, we believe that the time is here for a serious experimental investigation of {\sc supcode} to test its efficiency in producing error-resistant one- and two-qubit gates for spin qubit operations in semiconductor quantum dot systems.

Quantum dot spin qubits, particularly because of their scalability, are one of the primary candidates for the building blocks of a quantum computer. The noise-insensitive gates generated by {\sc supcode} help fill the need for precise and robust quantum control in these qubits. In this paper, we show how one may apply {\sc supcode} to produce noise-resistant single-qubit, two-qubit and multi-qubit operations. Not only do {\sc supcode} sequences respect all the fundamental experimental constraints associated with singlet-triplet qubits, but they also possess a remarkable robustness and flexibility when realistic, sample-dependent factors are taken into account. 
We therefore believe that a judicious use of {\sc supcode} is capable of bringing gate errors below the quantum error correction threshold, thus ushering in the possibility of fault-tolerant quantum computation in singlet-triplet semiconductor spin qubits.

This work is supported by LPS-CMTC and IARPA.


%

\end{document}